\documentclass[useAMS, usenatbib]{mn2e}
\usepackage{rotating, color, amssymb, epstopdf, mathtools, siunitx, url}
\usepackage{graphicx}

\def\fdg{\hbox{$.\!\!^\circ$}}

\def\USydney{$^{1}$}
\def\CAASTRO{$^{2}$}
\def\CASS{$^{3}$}
\def\UWisc{$^{4}$}
\def\ANU{$^{5}$}
\def\ASTRON{$^{6}$}
\def\Curtin{$^{7}$}
\def\SKASA{$^{8}$}
\def\CfA{$^{9}$}
\def\Rhodes{$^{10}$}
\def\ASU{$^{11}$}
\def\Haystack{$^{12}$}
\def\RRI{$^{13}$}
\def\UToronto{$^{14}$}
\def\MIT{$^{15}$}
\def\UW{$^{16}$}
\def\Victoria{$^{17}$}
\def\UMichigan{$^{18}$}
\def\Tata{$^{19}$}
\def\NRAO{$^{20}$}
\def\UMelbourne{$^{21}$}

\title[Ionospheric effects on time-domain astrophysics]{Quantifying ionospheric effects on time-domain astrophysics with the Murchison Widefield Array}

\author[Loi et al.]{
Shyeh Tjing Loi\USydney$^,$\CAASTRO\thanks{Email: sloi5113@uni.sydney.edu.au}
Tara Murphy\USydney$^,$\CAASTRO,
Martin E.~Bell\CAASTRO$^,$\CASS,
David L.~Kaplan\UWisc, \newauthor
Emil Lenc\USydney$^,$\CAASTRO,
Andr\'{e} R.~Offringa\CAASTRO$^,$\ANU$^,$\ASTRON,
Natasha Hurley-Walker\Curtin,\newauthor
G.~Bernardi\SKASA$^,$\CfA$^,$\Rhodes,
J.~D.~Bowman\ASU,
F.~Briggs\ANU,
R.~J.~Cappallo\Haystack, 
B.~E.~Corey\Haystack,\newauthor
A.~A.~Deshpande\RRI, 
D.~Emrich\Curtin,
B.~M.~Gaensler\USydney$^,$\CAASTRO$^,$\UToronto, 
R.~Goeke\MIT,\newauthor
L.~J.~Greenhill\CfA,
B.~J.~Hazelton\UW,
M.~Johnston-Hollitt\Victoria,
J.~C.~Kasper\CfA$^,$\UMichigan,\newauthor
E.~Kratzenberg\Haystack, 
C.~J.~Lonsdale\Haystack,
M.~J.~Lynch\Curtin, 
S.~R.~McWhirter\Haystack,\newauthor
D.~A.~Mitchell\CAASTRO$^,$\CASS, 
M.~F.~Morales\UW,
E.~Morgan\MIT, 
D.~Oberoi\Tata, 
S.~M.~Ord\CAASTRO$^,$\Curtin,\newauthor
T.~Prabu\RRI, 
A.~E.~E.~Rogers\Haystack,
A.~Roshi\NRAO, 
N.~Udaya~Shankar\RRI,\newauthor
K.~S.~Srivani\RRI, 
R.~Subrahmanyan\CAASTRO$^,$\RRI,
S.~J.~Tingay\CAASTRO$^,$\Curtin, 
M.~Waterson\ANU$^,$\Curtin,\newauthor
R.~B.~Wayth\CAASTRO$^,$\Curtin, 
R.~L.~Webster\CAASTRO$^,$\UMelbourne,
A.~R.~Whitney\Haystack, 
A.~Williams\Curtin,\newauthor
C.~L.~Williams\MIT\\
$^{1}$Sydney Institute for Astronomy, School of Physics, The University of Sydney, NSW 2006, Australia\\
$^{2}$ARC Centre of Excellence for All-sky Astrophysics (CAASTRO)\\
$^{3}$CSIRO Astronomy and Space Science (CASS), PO Box 76, Epping, NSW 1710, Australia\\
$^{4}$Department of Physics, University of Wisconsin--Milwaukee, Milwaukee, WI 53201, USA\\
$^{5}$Research School of Astronomy and Astrophysics, Australian National University, Canberra, ACT 2611, Australia\\
$^{6}$Netherlands Institute for Radio Astronomy (ASTRON), Postbus 2, 7990 AA Dwingeloo, The Netherlands\\
$^{7}$International Centre for Radio Astronomy Research, Curtin University, Bentley, WA 6102, Australia\\
$^{8}$Square Kilometre Array South Africa (SKA SA), Cape Town 7405, South Africa\\
$^{9}$Harvard-Smithsonian Center for Astrophysics, Cambridge, MA 02138, USA\\
$^{10}$Department of Physics and Electronics, Rhodes University, PO Box 94, Grahamstown, 6140, South Africa\\
$^{11}$School of Earth and Space Exploration, Arizona State University, Tempe, AZ 85287, USA\\
$^{12}$MIT Haystack Observatory, Westford, MA 01886, USA\\
$^{13}$Raman Research Institute, Bangalore 560080, India\\
$^{14}$Dunlap Institute for Astronomy and Astrophysics, The University of Toronto, ON M5S 3H4, Canada\\
$^{15}$Kavli Institute for Astrophysics and Space Research, Massachusetts Institute of Technology, Cambridge, MA 02139, USA\\
$^{16}$Department of Physics, University of Washington, Seattle, WA 98195, USA\\
$^{17}$School of Chemical \& Physical Sciences, Victoria University of Wellington, Wellington 6140, New Zealand\\
$^{18}$Department of Atmospheric, Oceanic and Space Sciences, University of Michigan, Ann Arbor, MI 48109, USA\\
$^{19}$National Centre for Radio Astrophysics, Tata Institute for Fundamental Research, Pune 411007, India\\
$^{20}$National Radio Astronomy Observatory, Charlottesville and Greenbank, USA\\
$^{21}$School of Physics, The University of Melbourne, Parkville, VIC 3010, Australia}
\date{Last compiled: \today}

\begin{document}
\maketitle

\clearpage
\begin{abstract}
Refraction and diffraction of incoming radio waves by the ionosphere induce time variability in the angular positions, peak amplitudes and shapes of radio sources, potentially complicating the automated cross-matching and identification of transient and variable radio sources. In this work, we empirically assess the effects of the ionosphere on data taken by the Murchison Widefield Array (MWA) radio telescope. We directly examine 51 hours of data observed over 10 nights under quiet geomagnetic conditions (global storm index $K_p < 2$), analysing the behaviour of short-timescale angular position and peak flux density variations of around ten thousand unresolved sources. We find that while much of the variation in angular position can be attributed to ionospheric refraction, the characteristic displacements (10--20\,arcsec) at 154\,MHz are small enough that search radii of 1--2\,arcmin should be sufficient for cross-matching under typical conditions. By examining bulk trends in amplitude variability, we place upper limits on the modulation index associated with ionospheric scintillation of 1--3\% for the various nights. For sources fainter than $\sim$1\,Jy, this variation is below the image noise at typical MWA sensitivities. Our results demonstrate that the ionosphere is not a significant impediment to the goals of time-domain science with the MWA at 154\,MHz.
\end{abstract}


\section{Introduction}
The fast survey capabilities of next-generation radio telescopes such as the Murchison Widefield Array \citep[MWA;][]{Lonsdale2009_mn2e, Bowman2013_mn2e, Tingay2013_mn2e}, the Australian Square Kilometre Array Pathfinder \citep[ASKAP;][]{Johnston2008_mn2e, Hotan2014_mn2e} and the Low Frequency Array \citep[LOFAR;][]{vanHaarlem2013} greatly facilitate the study of transient and variable radio sources. The primary observables in such surveys are sky positions and flux densities, while secondary observables may include angular source sizes and shapes, polarisation fractions, polarisation vectors and rotation measures. 

Many astrophysical phenomena, including gamma-ray bursts, stellar flares, pulsars, exoplanets, black hole jet launches, supernovae and fast radio bursts produce transient and variable radio emission on a wide range of time scales, from milliseconds to days. Locating and identifying the progenitors of these events has been the focus of many recent blind surveys making use of new low-frequency telescopes \citep{Jaeger2012, Bell2014_mn2e, Carbone2014_mn2e, Obenberger2015_mn2e} and also upcoming surveys at higher frequencies \citep{Murphy2013_mn2e}. Given the high data rate from these new instruments \citep{Norris2010}, the extraction, cross-matching, light-curve generation and classification of sources must be conducted in an automated fashion \citep{Murphy2013_mn2e, Lo2014, Swinbank2015_mn2e}. An important aspect of these automated pipelines is that major sources of error and extrinsic/foreground variability are understood and accounted for to prevent source misclassifications. The recent discovery of radio emission from fireballs \citep{Obenberger2014} exemplifies a previously unknown terrestrial source of transient radio emission that may confuse searches for celestial transients.

One important contributor to extrinsic variability, especially at frequencies less than $\sim$1\,GHz, is the distortion of phase fronts due to propagation through the Earth's ionosphere. The ionosphere is the ionised component of the Earth's atmosphere. It extends between altitudes of roughly 50--1000\,km, with the electron density peaking near 300--400\,km altitude at night \citep{Luhmann1995, Solomon2010}. At low radio frequencies, spatially-varying electron densities in the ionosphere induce variations in the angular position and flux density of radio sources through refraction and diffraction of propagating radio waves \citep{Bougeret1981, Jacobson1992, Kassim2007}. Position shifts can diminish the accuracy of cross-matching between different epochs of observation, and propagation-induced amplitude variability can result in celestial source misclassifications. Smearing and angular broadening can cause otherwise point-like sources to become distorted, and this can pose difficulties for deconvolution and source identification/extraction. Quantifying these effects is important both for the design and optimisation of algorithms that remove ionospheric distortions from the data \citep{Cotton2004, Intema2009}, and also to obtain realistic error bars on observables for time-domain astrophysics \citep[cf.][]{Bell2014_mn2e}.

A number of studies have been aimed at quantifying the effects of the ionosphere on radio astronomical observations at low radio frequencies. These include the theoretical work of \citet{Hewish1951} and \citet{Meyer-Vernet1980}, and observational work with the Nan\c{c}ay radio interferometer \citep{Bougeret1981, Mercier1986}, the Westerbork Radio Synthesis Telescope \citep{Spoelstra1983, Spoelstra1984, Spoelstra1985, Spoelstra1997}, the Los Alamos radio interferometer \citep{Jacobson1996, Hoogeveen1997a}, and the Very Large Array (VLA) radio interferometer \citep{Jacobson1992a, Jacobson1992, Jacobson1993, Hoogeveen1997, Coker2009, Cohen2009, Helmboldt2012, Helmboldt2012a, Helmboldt2012b, Helmboldt2014a}. These have established that radio telescopes are highly sensitive to and can even be used to characterise ionospheric perturbations and irregularities ordinarily studied through other means, by isolating propagation-induced fluctuations in the amplitudes, phases and angular positions of radio sources. The effects of electron density fluctuations occurring on timescales of several minutes and with spatial scales of $\sim$10\,km have been observed in interferometric data. This is important because the characteristic spatiotemporal scales of ionospheric behaviour determine the spatiotemporal scales on which calibration is required, and ultimately the precision to which astronomical observables can be measured \citep{Braun2013}.

The types and occurrence rates of ionospheric fluctuations depend on geographic location, tropospheric weather and geomagnetic conditions \citep{Schunk2009}. Recent work by \citet{Datta2014} and \citet{Arora2015_mn2e} has explored the possibility of using Global Positioning System (GPS) satellite measurements to calibrate for ionospheric effects in radio interferometric data. However, the existence of a substantial population of irregularities on length scales below the resolution of global GPS maps \citep[$\sim$100\,km;][]{Rideout2006} implies that GPS measurements alone cannot fully predict the extent to which the ionosphere affects radio observations \citep{Loi2015a_mn2e}. Furthermore, the coherence properties of ionospheric irregularities of a given scale size depend on the configuration of the array, with decoherence and scintillation effects more severe on longer baselines \citep{Lonsdale2005, Kassim2007, Intema2009}. Although general statements can be made based on physical arguments and established knowledge (see \S\ref{sec:effects}), a direct examination of the data is the most straightforward means of assessing the severity of ionospheric effects for a given radio interferometer and geographical site.

This work aims to empirically establish the effects of the ionosphere on radio astronomy conducted with the MWA, focusing on its impacts on time-domain science, for which the two most important observables are angular position and flux density. We directly examine the statistics of those observables in 10 nights of data collected as part of the MWA Transients Survey (MWATS; PI: M.~E.~Bell), a general-purpose, low-frequency (154\,MHz) radio survey looking for transient and variable sources. We begin in \S\ref{sec:effects} with an overview of how the ionosphere affects radio interferometric measurements. We describe our observations and analysis approach in \S\ref{sec:obs}, and present the results in \S\ref{sec:results}. We interpret the results and discuss their implications in \S\ref{sec:discussion}. 

Angular broadening and shape distortion effects are not considered here, although they may diminish the performance of astronomical source-finding algorithms (by making the source more difficult to identify, or increasing the measurement error in flux density and/or position). As argued in \S\ref{sec:overviewposition}, distortion and broadening effects are expected to be secondary to angular position shifts of whole sources, given the compact physical size of the MWA, for sufficiently short integrations.

\section{Overview of Ionospheric Effects}\label{sec:effects}
Electron densities in the ionosphere are commonly described in terms of the total electron content (TEC; i.e.~the electron column density $\int n_e \,\mathrm{d}\ell$, where $n_e$ is the local electron number density and $\ell$ parametrises distance along the line of sight), whose units are electrons\,m$^{-2}$, commonly expressed in TEC units (TECU; 1\,TECU $\equiv 10^{16}$\,electrons\,m$^{-2}$). Radio waves from a given source arrive at the different receivers of an interferometer having traversed different paths through the ionosphere, acquiring different phase delays if the density is non-uniform. An initially planar set of wavefronts thus becomes distorted. This can cause the source to appear shifted from its true position, its peak amplitude to vary and/or its shape to distort \citep{Smith1952, Spoelstra1997, Intema2009}. Multi-path propagation through a highly irregular plasma can lead to self-interference of coherent phase fronts, causing a diffraction pattern to develop and giving rise to scintillation \citep{Meyer-Vernet1980}. Phase delays associated with water vapour in the troposphere, although significant at high ($\gtrsim$1\,GHz) frequencies, are $\sim$100 times smaller than those of the ionosphere at the low frequencies of the MWA \citep{Thompson2001}. Furthermore, the aridity of the site ensures that tropospheric activity is relatively benign. The tropospheric contribution can therefore be largely ignored, and we consider only ionospheric effects here.

\subsection{Angular position}\label{sec:overviewposition}
If the TEC along the lines of sight from all receivers to a given source is the same, then the only effect (ignoring those related to polarisation and frequency dispersion, which we do not consider here) is to add a constant phase to all antennas. A constant absolute phase is not measurable by correlation interferometry, and so the recorded visibilities are identical to what they would be if there were no ionosphere. An angularly uniform TEC screen therefore has no effect on radio observations; distortions only arise from density inhomogeneities.

For radio waves propagating well above the local electron plasma frequency $\omega_p = (e^2 n_e/\epsilon_0 m_e)^{1/2}$, the refractive index $n$ is given by $n \approx 1 - \omega_p^2/2\omega^2$, where $\omega \gg \omega_p$ is the angular frequency of the radio waves, $\epsilon_0$ is the vacuum permittivity, and $e$ and $m_e$ are the electron charge and mass. Consider the lines of sight $\ell_1$ and $\ell_2$ from two separate interferometer elements to a given radio source, where the elements are spaced by a distance $D$. The optical path difference between $\ell_1$ and $\ell_2$ is
\begin{align}
  \mathrm{OPD} = \int_{\ell_1} n(x) \,dx - \int_{\ell_2} n(x) \,dx = -\frac{1}{2} \frac{e^2}{\omega^2 \epsilon_0 m_e} \Delta \mathrm{TEC} \:, \label{eq:OPD}
\end{align}
where $\Delta \mathrm{TEC} = \mathrm{TEC}(\ell_1)- \mathrm{TEC}(\ell_2)$. If $\Delta \mathrm{TEC} \neq 0$, then the wavefront appears to be tilted with respect to the original and this causes the apparent position of the source to shift \citep{Smith1952, Thompson2001}. The phase difference between the antennas is
\begin{align}
  \Delta \phi = \phi(\ell_1) - \phi(\ell_2) = \frac{\omega \mathrm{OPD}}{c} = -\frac{e^2}{4\pi \epsilon_0 c m_e} \frac{1}{\nu} \Delta \mathrm{TEC} \:, \label{eq:phasediff}
\end{align}
where $\nu = \omega/2\pi$ is the observing frequency. In the small-angle approximation and assuming parallel lines of sight, the resulting angular shift in position is
\begin{align}
  \Delta \theta \approx \frac{\mathrm{OPD}}{D} = -\frac{1}{8\pi^2} \frac{e^2}{\epsilon_0 m_e} \frac{1}{\nu^2} \frac{\Delta \mathrm{TEC}}{D} \:. \label{eq:posshift}
\end{align}
The negative sign in Equation (\ref{eq:posshift}) indicates that sources refract in the direction of decreasing TEC. The $\nu^{-2}$ dependence implies that for a given $\Delta \mathrm{TEC}/D$, the refractive shift is larger at lower frequencies.

If the collection of sight lines from all antennas to a given source passes through a patch of the ionosphere whose TEC distribution is well described by a linear ramp (insignificant spatial curvature), then $\Delta \mathrm{TEC}$ is linearly proportional to $D$. Equation (\ref{eq:posshift}) then implies that $\Delta \theta$ will be the same on all baselines. Substituting values for the various physical constants, this yields
\begin{align}
  \Delta \theta \approx -\frac{40.3}{\nu^2} \nabla_\perp \mathrm{TEC} \label{eq:posshift2}
\end{align}
as the angular offset of the radio source in an image synthesised using the array, where $\Delta \theta$ is in radians, $\nu$ is in Hz and $\nabla_\perp \mathrm{TEC}$ (the transverse gradient of the TEC) is in electrons\,m$^{-3}$. At mid-latitudes where the MWA is sited, medium-scale travelling ionospheric disturbances (TIDs) driven by atmospheric waves are the most familiar type of wavelike perturbation \citep{Thompson2001}. These have associated $\nabla_\perp \mathrm{TEC}$ values of $10^{10}$--$10^{11}$\,electrons\,m$^{-3}$ \citep{Jacobson1992a, Jacobson1992, Dymond2011_mn2e}, which translates to $\Delta \theta \sim 10$\,arcsec at $\nu$ = 154\,MHz. 

As long as the length scales of the perturbation are much larger than the physical size of the array, irrespective of the field-of-view (FoV), and $\nabla_\perp \mathrm{TEC}$ is roughly constant within an integration time, the perturbed wavefront remains approximately planar over the array and distortions are primarily in the form of angular position shifts with minimal changes to source shape or amplitude. The physical size of the MWA, which has a longest baseline of $\sim$3\,km \citep[comparable to the mean free path of neutral particles\footnote{Mostly N$_2$, O$_2$, O and He at the relevant altitudes.} in the thermosphere;][]{Jacchia1977} is small compared to the wavelengths of TIDs, which can be 100--1000\,km \citep{Hunsucker1982}. This suggests that neutral gas motions at ionospheric heights, including the atmospheric waves that drive TIDs, are locally expected to produce only simple position shifts of point sources in MWA images. However, the energy of these disturbances may potentially drive cascades of smaller-scale irregularities through various plasma instabilities \citep{Whitehead1971, Klostermeyer1978, Rottger1978}. Plasma turbulence occurring on scales below 3\,km down to the ion gyroradius \citep[$\sim$5\,m in the thermosphere;][]{Woodman1978} may produce higher-order wavefront perturbations over the MWA \citep{Booker1979}, leading to source broadening and shape distortions.

\subsection{Flux density}\label{sec:overviewflux}
While position shifts are governed primarily by the first spatial derivative of the TEC, scintillation and angular broadening effects are related to higher-order derivatives \citep{Lee1975, Meyer-Vernet1980}. The scattering and diffusion of ray paths upon propagation through an inhomogeneous medium causes focusing and defocusing of radio waves. Curvature of an initially planar wavefront causes otherwise point-like sources to acquire a finite coherence length, leading to an increase in apparent width and a decrease in peak amplitude. Temporal variations in the signal may come from intrinsic time variability in the plasma, and/or relative motion between the plasma and the lines of sight to celestial sources. This relative drift arises both from the rotation of the Earth ($\sim$20\,m\,s$^{-1}$ drift of sight lines through the ionosphere at the latitude of the MWA, assuming an ionospheric altitude of 300\,km), and also neutral winds causing the bulk plasma to drift at 10--100\,m\,s$^{-1}$ with respect to the ground \citep{Davies1973}. Small systematic deviations in flux density may also arise from absorption due to collisions between electrons and ions/neutrals, this being $\sim$1\% in the daytime and $\sim$0.1\% at night for typical MWA operating frequencies \citep{Thompson2001}.

For perturbations to the phase fronts by an amount less than $\sim$1\,radian within a patch of some critical length $L$, amplitude distortions are small (fractional variations much less than unity). This is the regime of weak scintillation, inhabited by the ionosphere most of the time for frequencies around 100\,MHz \citep{Narayan1992}. The timescale of amplitude variability is $L/v$, where $v$ is the relative drift velocity between the ionosphere and the sight lines \citep{Narayan1989}. The coherence length $L$ is given by the larger of either the baseline length or the Fresnel scale $r_F = \sqrt{\lambda z}$, the latter of which describes the size of the region on the ionosphere from which Huygen wavelets arrive at a point on the ground approximately in phase \citep{Cronyn1972}. Here $\lambda$ is the observing wavelength and $z$ is the altitude of the ionosphere. Baseline lengths for the MWA (several hundred metres) are comparable to or below the Fresnel scale at 154\,MHz ($\sim$1\,km), implying a characteristic timescale of 10--100\,s for amplitude variability due to the ionosphere.

If deflection angles are sufficiently large, rays from a given celestial source intersect before they reach the ground, producing a fully-developed Fraunhofer diffraction pattern that gives rise to large and rapid amplitude fluctuations (fractional variations of order unity). This regime is known as strong scintillation, and occurs when there are significant phase fluctuations on spatial scales below $r_F$. If such irregularities are produced by a turbulent cascade driven by motions of the neutral atmosphere, then we might also expect such events to be accompanied by significant position fluctuations induced by longer-wavelength modes within the inertial range (since outer scales would exceed the neutral mean free path of 3\,km, and thereby the size of the MWA). Since $r_F \propto 1/\sqrt{\nu}$ and $\Delta \phi \propto 1/\nu$, when $\nu$ is small irregularities of a given scale size are both associated with larger phase fluctuations and have a greater chance of falling below $r_F$. Scintillation events are therefore more likely to occur at low frequencies.

A source whose angular size exceeds $L/z$ will not scintillate, because the diffraction patterns associated with different parts of the source will blend together and average out. For the MWA at 154\,MHz, the critical angular diameter below which a source can scintillate, given by the angle subtended by the Fresnel scale at ionospheric heights, is about 10\,arcmin. Such a source would be resolved in the images (synthesised beamwidth $\sim$2\,arcmin at 154\,MHz). The vast majority of radio sources in the sky are not resolved by the MWA, and so ionospheric scintillation events, when they occur, would be expected to affect a large number of sources in the data. This contrasts interplanetary and interstellar scintillation, which are only expected to affect the most compact sources \citep{Thompson2001, Kaplan2015_mn2e}.

One way to distinguish ionosphere-induced variability from intrinsic variability is to check whether neighbouring sources are varying with a similar modulation index on a similar timescale. However, if it so happens that a highly localised patch of irregularities conspires to affect only an isolated source (thus mimicking intrinsic variability), it is unlikely that this patch will remain in front of the source for very long. Because of the rotation of the Earth, in-situ stationary ionospheric irregularities will drift with respect to celestial sight lines at a rate of about 0.2\,deg\,min$^{-1}$. Given an average source spacing of 0.5--1$^\circ$ for a 2-min integration with the MWA, it will only take a few minutes a given source to drift out of its patch of ionosphere and into its neighbour's. Variability of an isolated source that continues for more than several minutes is therefore more likely to be intrinsic (or perhaps of interplanetary/interstellar origin) than ionospheric.

Unlike position offsets, which are controlled by the largest-scale density structures, amplitude variations are governed by irregularities on the smallest scales. Large-scale density variations associated with TIDs and whistler ducts, the two most common types of structures appearing in MWA data \citep{Loi2015a_mn2e}, produce only negligible flux density changes: for the relatively extreme TEC gradients measured by \citet{Loi2015_mn2e} for whistler ducts, with angular deflections of $\Delta \theta \sim$1\,arcmin over $L \sim 10$\,km transverse scales, the associated fractional amplitude variations are estimated to be\footnote{We arrived at Equation (\ref{eq:weakfluxvarn}) under simple geometric optics considerations, which are approximately valid under conditions of weak scintillation (no interference effects).}
\begin{align}
  \frac{\Delta S}{S} \approx \frac{z \Delta \theta}{L} \sim 1\% \:, \label{eq:weakfluxvarn}
\end{align}
where $S$ is the flux density. For TIDs, which have $\Delta \theta \sim 10$\,arcsec and $L \sim 100$\,km, we find that $\Delta S/S \sim 0.01\%$. This is well below instrument sensitivities for 2-min integrations ($\sim$50\,mJy; this has contributions from thermal noise, and also classical and sidelobe confusion) for typical source brightnesses ($\sim$0.5\,Jy) at 154\,MHz.

A more sophisticated treatment is required for small-scale irregularities. Theoretical work considering the effects of scintillation on interferometric visibilities has been conducted by a number of authors, under various assumptions and approximations \citep{Cronyn1972, Goodman1989, Koopmans2010, Vedantham2014}. A thin phase-screen model is often used, and the spectrum of small-scale fluctuations assumed to be a power law \citep{Rino1982, Wheelon2003}. The results have been shown to depend on many quantities, including the strength, spectral index and inner and outer scales for the phase fluctuation spectrum, the channel width, integration time and frequency of the observations, the altitude and thickness of the phase screen, and the FoV and geometry of the array. Compared to position offsets, which depend on just two free parameters ($\nu$ and $\nabla_\perp \mathrm{TEC}$), it is far less straightforward to estimate the variability of peak flux densities in the final synthesised images. The recent theoretical work of \citet{Vedantham2014} for realistic turbulence parameters suggests that ionospheric scintillation may be a substantial source of noise variance in the amplitudes and phases of visibilities recorded by widefield, low-frequency arrays. Here we empirically assess scintillation effects on source flux densities for the case of the MWA at 154\,MHz.

\section{Observations, Data Reduction \& Analysis}\label{sec:obs}
Part of the usual calibration process of radio interferometric data is a removal of ionospheric effects. For conventional instruments, which have narrow fields of view and operate at higher frequencies, the ionosphere can be assumed to be locally uniform within the FoV of each antenna. In this situation, a single phase term for each antenna is sufficient for capturing the effects of the ionosphere. Calibration algorithms such as phase referencing \citep{Fomalont1999} and self-calibration \citep{Pearson1984, Cornwell1999}, which solve for a single phase correction term per antenna, are effective at removing ionospheric phases from the data. 

In contrast, the widefield nature of next-generation instruments implies that each antenna sees a large patch of the ionosphere, and the assumption of local uniformity breaks down \citep{Cotton2004, Lonsdale2005, Intema2009_thesis}. The ionospheric phase is now a direction-dependent function for each antenna, and this cannot be removed using conventional calibration algorithms, which neglect direction dependencies. In this study, we analyse data for which only time-independent and direction-independent corrections to the antenna phases have been applied. The wide FoV of the MWA, combined with its compact size, imply that antennas see almost identical phase screens and so differences in the ionospheric phase between antennas are expected to be small. The calibration process to which the data were subjected is therefore only effective for removing instrumental phase errors (e.g.~due to slow temperature-induced changes in cable lengths), and leaves the ionospheric phase (dominated by rapid time- and direction-dependent rather than antenna-dependent variations) largely untouched.

\subsection{Observing Strategy}
The data were recorded over a 30.72\,MHz band centred at 154\,MHz. Data collection began in mid-2013 and took place one night each month. During an observing run, which typically lasted for 9--10\,hr, the telescope was pointed along the meridian at three different declinations, centred at zenith and zenith angles of $\sim$30$^\circ$ ($\delta = -55\fdg0, -26\fdg7, +1\fdg6$). A series of 2-min snapshots were obtained at each declination in turn, the cycle repeating every 6\,min. The drift of the sky during each 2-min scan was accounted for during imaging by fixing the phase centre for each snapshot to be at a certain RA/Dec. We have sub-divided the data by month and declination band, such that each dataset corresponds to a fixed pointing direction (in Az/El) and contains snapshots from a single night. The naming convention here is ``year month declination'', where for example ``2013\,Sep\,$-$55'' refers to the dataset obtained in September of 2013, with snapshots centred at $\delta = -55\fdg0$ on the meridian.

This study analyses all MWATS datasets currently available that have passed quality control, namely those for which post-calibration gain fluctuations lie within acceptable tolerances. Although they were not selected based on any measure of ionospheric, geomagnetic, tropospheric or solar activity, we note that all MWATS observations were done at night and happened to be conducted under quiet geomagnetic conditions ($K_p$ index $\lesssim 2$). Our results should therefore be representative of typical quiet nighttime conditions above the observatory. It should be emphasised that these results do not apply to other times of the day, since the electron content undergoes large diurnal modulations, or observatories at other locations, because of the strong dependence on geomagnetic latitude. Ionospheric irregularities are much more prevalent at equatorial and high latitudes, whereas mid-latitude sites such as where the MWA is situated are known to experience lower levels of activity \citep{Fejer1980}.

There are a total of 20 datasets accounting for 10 different nights and a total of about 51 hours of observation, spread over a period of slightly more than a year between 2013 and 2014 (observing dates/times and central coordinates are listed in Table \ref{tab:datasetparams}). We make direct use of the standard survey images with no special reprocessing. Note that MWATS is intended for general-purpose astrophysical transient and variable searches, and is not designed for characterisation of the ionosphere. However, given that the aim of our study is to examine the variability that manifests at the time and frequency resolution of MWATS (and not, for example, to constrain the structure of the ionosphere), we consider the data sufficient for this purpose.

\begin{table*}
  \centering
  \begin{minipage}{20cm}
  \caption{Parameters for each dataset.}
  \begin{tabular}{lccS[table-format=3.2]lll} \hline
    Dataset & UTC start/end & AWST\footnote{AWST = Australian Western Standard Time (UTC +8)} start/end & \multicolumn{1}{c}{LST start/} & Central RA/Dec & Central Az/El & Calibrator \\
     & & & \multicolumn{1}{c}{end (hrs)} & & \\ \hline
    2013\,Sep\,$-55$ & 2013-09-16 13:30:39 & 2013-09-16 21:30:40 & 21.0 & $12^\circ$, $-55^\circ$ & 180\fdg0, 61\fdg7 & PKS 2356$-$61 \\
    & 2013-09-16 21:24:39 & 2013-09-17 05:24:40 & 4.9 & & \\
    2013\,Oct\,$-26$ & 2013-10-16 13:40:39 & 2013-10-16 21:40:40 & 23.1 & $44^\circ$, $-27^\circ$ & 0\fdg0, 90\fdg0 & 3C444 \\
    & 2013-10-16 21:34:39 & 2013-10-17 05:34:40 & 7.1 & & \\
    2013\,Dec\,$+1.6$ & 2013-12-06 13:51:28 & 2013-12-06 21:51:28 & 2.7 & $87^\circ$, $+2^\circ$ & 0\fdg0, 61\fdg7 & Hydra A \\
    & 2013-12-06 20:09:28 & 2013-12-07 04:09:28 & 9.0 & & \\
    2013\,Dec\,$-26$ & 2013-12-06 13:55:27 & 2013-12-06 21:55:28 & 2.7 & $88^\circ$, $-27^\circ$ & 0\fdg0, 90\fdg0 & Hydra A \\
    & 2013-12-06 20:07:27 & 2013-12-07 04:07:28 & 9.0 & & \\
    2013\,Dec\,$-55$ & 2013-12-06 13:53:28 & 2013-12-06 21:53:28 & 2.7 & $86^\circ$, $-55^\circ$ & 180\fdg0, 61\fdg7 & Pictor A \\
    & 2013-12-06 20:05:28 & 2013-12-07 04:05:28 & 8.9 & & \\
    2014\,Mar\,$+1.6$ & 2014-03-06 11:20:48 & 2014-03-06 19:20:48 & 6.1 & $161^\circ$, $+2^\circ$ & 0\fdg0, 61\fdg7 & Hercules A \\
    & 2014-03-06 21:00:48 & 2014-03-07 05:00:48 & 15.8 & & \\
    2014\,Mar\,$-26$ & 2014-03-03 11:32:32 & 2014-03-03 19:32:32 & 6.1 & $161^\circ$, $-27^\circ$ & 0\fdg0, 90\fdg0 & Hydra A \\
    & 2014-03-03 21:12:32 & 2014-03-04 05:12:32 & 15.8 & & \\
    2014\,Mar\,$-55$ & 2014-03-17 11:07:28 & 2014-03-17 19:07:28 & 6.6 & $143^\circ$, $-55^\circ$ & 180\fdg0, 61\fdg7 & Pictor A \\
   & 2014-03-17 20:47:28 & 2014-03-18 04:47:28 & 16.3 & & \\
    2014\,Apr\,$+1.6$ & 2014-04-28 10:52:00 & 2014-04-28 18:52:00 & 9.1 & $210^\circ$, $+2^\circ$ & 0\fdg0, 61\fdg7 & Hercules A \\
    & 2014-04-28 20:46:00 & 2014-04-29 04:46:00 & 19.0 & & \\
    2014\,Apr\,$-26$ & 2014-04-28 10:50:00 & 2014-04-28 18:50:00 & 9.0 & $209^\circ$, $-27^\circ$ & 0\fdg0, 90\fdg0 & Hydra A \\
    & 2014-04-28 20:43:52 & 2014-04-29 04:43:52 & 19.0 & & \\
    2014\,Apr\,$-55$ & 2014-04-28 10:48:00 & 2014-04-28 18:48:00 & 9.0 & $209^\circ$, $-55^\circ$ & 180\fdg0, 61\fdg7 & Virgo A \\
   & 2014-04-28 20:42:00 & 2014-04-29 04:42:00 & 19.0 & & \\
    2014\,Jul\,$+1.6$ & 2014-07-15 10:46:24 & 2014-07-15 18:46:24 & 14.1 & $285^\circ$, $+2^\circ$ & 0\fdg0, 61\fdg7 & Hercules A \\
    & 2014-07-15 20:40:24 & 2014-07-16 04:40:24 & 0.0 & & \\
    2014\,Jul\,$-26$ & 2014-07-15 10:44:24 & 2014-07-15 18:44:24 & 14.1 & $285^\circ$, $-27^\circ$ & 0\fdg0, 90\fdg0 & 3C444 \\
    & 2014-07-15 20:38:24 & 2014-07-16 04:38:24 & 0.0 & & \\
    2014\,Jul\,$-55$ & 2014-07-15 10:42:24 & 2014-07-15 18:42:24 & 14.0 & $284^\circ$, $-55^\circ$ & 180\fdg0, 61\fdg7 & 3C444 \\
    & 2014-07-15 20:36:24 & 2014-07-16 04:36:24 & 0.0 & & \\
    2014\,Aug\,$+1.6$ & 2014-08-26 11:00:48 & 2014-08-26 19:00:48 & 17.1 & $330^\circ$, $+2^\circ$ & 0\fdg0, 61\fdg7 & Hercules A \\
    & 2014-08-26 20:54:48 & 2014-08-27 04:54:48 & 3.0 & & \\
    2014\,Aug\,$-26$ & 2014-08-26 10:58:48 & 2014-08-26 18:58:48 & 17.1 & $330^\circ$, $-27^\circ$ & 0\fdg0, 90\fdg0 & 3C444 \\
   & 2014-08-26 20:52:48 & 2014-08-27 04:52:48 & 3.0 & & \\
    2014\,Aug\,$-55$ & 2014-08-26 10:56:48 & 2014-08-26 18:56:48 & 17.1 & $329^\circ$, $-55^\circ$ & 180\fdg0, 61\fdg7 & PKS 2356$-$61 \\
    & 2014-08-26 20:50:48 & 2014-08-27 04:50:48 & 3.0 & & \\
    2014\,Oct\,$+1.6$ & 2014-10-07 11:15:12 & 2014-10-07 19:15:12 & 20.1 & $15^\circ$, $+2^\circ$ & 0\fdg0, 61\fdg7 & 3C444 \\
    & 2014-10-07 21:09:12 & 2014-10-08 05:09:12 & 6.0 & & \\
    2014\,Oct\,$-26$ & 2014-10-07 11:13:12 & 2014-10-07 19:13:12 & 20.1 & $15^\circ$, $-27^\circ$ & 0\fdg0, 90\fdg0 & 3C444 \\
   & 2014-10-07 21:07:12 & 2014-10-08 05:07:12 & 6.0 & & \\
    2014\,Oct\,$-55$ & 2014-10-07 11:11:12 & 2014-10-07 19:11:12 & 20.0 & $14^\circ$, $-55^\circ$ & 180\fdg0, 61\fdg7 & Pictor A \\
    & 2014-10-07 21:05:12 & 2014-10-08 05:05:12 & 6.0 & & \\ \hline
  \end{tabular}\\
  \label{tab:datasetparams}
  \end{minipage}
\end{table*}

\subsection{MWA Data Reduction}
To calibrate each dataset, we observed a bright source with well-modelled emission (a calibrator) for two minutes at the phase centre of the instrument. We chose the nearest available calibrator in declination to the observations. A reference image, taken from another low-frequency instrument, was used as the starting point for calibration. Time-independent, frequency-dependent phase and ampltude calibration solutions were derived based on a frequency-adjusted model of each calibrator. Table \ref{tab:datasetparams} lists the calibrators used for each dataset.


We used a pre-processing algorithm to flag radio-frequency interference (RFI), average the data and then convert it into CASA\footnote{\url{http://casa.nrao.edu}} (version 4.2) measurement set format. RFI flagging was achieved using the algorithm \textsc{aoflagger} \citep{Offringa2012}. The calibration solutions were then applied to the visibilities, which were imaged and deconvolved using the \textsc{wsclean} algorithm \citep{Offringa2014}. An image size of $3072 \times 3072$ pixels was used with a pixel size of 45\,arcsec. Stokes I images were formed using a Briggs weighting of $-1$, giving a result closer to uniform than natural weighting. The \textsc{wsclean} algorithm achieves this by forming a complex $2 \times 2$ Jones matrix $I$ for each image pixel. The images were restored using a circular Gaussian of width 130\,arcsec, which roughly describes the shape of the MWA synthesised beam at zenith, and this was elongated appropriately off-zenith to account for the foreshortening of projected instrument baselines. Primary beam correction was performed by inverting the beam voltage matrix $B$ and computing $B^{-1} I B^{*-1}$, where $^*$ denotes the conjugate transpose \citep[details in][]{Offringa2014}. The resulting Stokes I snapshots were then used in subsequent analysis. It is to be noted that residual errors exist in the analytical beam model; these are discussed further in \S\ref{sec:methodflux}.

\subsection{Source Extraction and Selection}
We extracted sources from the snapshot images using the source finder \textsc{Aegean} \citep{Hancock2012}. This fits a 2D Gaussian to each source it detects above a certain noise threshold. The outputs from \textsc{Aegean} relevant for this analysis are the best-fit angular position, peak flux density, Gaussian shape parameters and their associated errors. Among the extracted sources, we excluded those with negative peak flux densities, those with position errors exceeding 1\,arcmin (as quoted by \textsc{Aegean}, and corresponding to the positional uncertainty on a source with S/N ratio $\sim$1), and those for which there were fewer pixels above 5$\sigma$ than the number of parameters to be fitted. The purpose of these restrictions was to remove the poorer-quality fits.

We then cross-matched the remainder with either the NRAO VLA Sky Survey (NVSS) catalogue \citep{Condon1998} or the Sydney University Molonglo Sky Survey (SUMSS) catalogue \citep{Mauch2003}, depending on the declination range of the dataset\footnote{NVSS covers $\delta > -40^\circ$, while SUMSS covers $\delta < -30^\circ$. We cross-matched datasets having $\delta = -26\fdg7, +1\fdg6$ with NVSS and those having $\delta = -55\fdg0$ with SUMSS.}. The cross-matching radius we used for most datasets was 1.2\,arcmin. However, for the datasets 2014\,Aug\,$-26$ and 2014\,Aug$-55$, where extreme position offsets were detected (of order the synthesised beam or larger), we had to increase the cross-matching radius to 3.6\,arcmin to obtain a cross-matching efficiency (fraction of \textsc{Aegean} sources with a match in the external catalogue) comparable to the other datasets ($\sim$95\%).

To reduce the fraction of spurious sources (which may be noise spikes or residual sidelobes), we discarded all sources that did not have a match in the external catalogues. We also placed a restriction on the minimum number of snapshots a source had to appear in (on a dataset-by-dataset basis) to be considered for subsequent analysis. The value of the restriction was 20 snapshots (roughly 75\% of the maximum number of snapshots that a source at the central declination will appear in) for most datasets, but we lowered this to 10 for several smaller datasets that had very few sources appearing in more than 20 snapshots. This was chosen as a compromise between excluding sources whose true positions were less well represented by the time-averaged positions (see later in \S\ref{sec:methodpos}), and retaining a sufficiently large sample for statistical analysis. Note that many sources appear in more than 20 snapshots, particularly for the southern MWATS pointing where 30--40\% of sources are present in 30 or more snapshots. The value of the cut on the number of snapshots and the resulting numbers of sources analysed are stated in Table \ref{tab:sourcenumbers}, which also lists the total number of snapshots and source occurrences for each dataset. Note that this filtering implies that the resulting source sample is not complete. Rather, it is intended to be a reliable sample of sources that can be used to probe ionospheric fluctuations.

\begin{table*}
  \begin{minipage}{15cm}
  \centering
  \caption{Snapshot and source numbers for each dataset. The third column lists the minimum number of snapshots a source in a particular dataset was required to appear in to be included for further analysis. The fourth column refers to the number of distinct SUMSS or NVSS sources matched to the candidate sources extracted by \textsc{Aegean}, subject to their appearing in at least the number of snapshots listed in the third column. The last column refers to the total number of appearances of these sources in any snapshot.}
  \begin{tabular}{lrrrr} \hline
    Dataset & Snapshots & Snapshot cut & Sources & Source occurrences \\ \hline
    2013\,Sep\,$-55$ & 76 & 20 & 2923 & 75514 \\
    2013\,Oct\,$-26$ & 76 & 20 & 2894 & 67293 \\
    2013\,Dec\,$+1.6$ & 56 & 10 & 2780 & 38240 \\
    2013\,Dec\,$-26$ & 58 & 20 & 860 & 19851 \\
    2013\,Dec\,$-55$ & 60 & 20 & 1239 & 33800 \\
    2014\,Mar\,$+1.6$ & 55 & 10 & 2431 & 28350 \\
    2014\,Mar\,$-26$ & 54 & 10 & 5069 & 63819 \\
    2014\,Mar\,$-55$ & 52 & 10 & 2142 & 37845 \\
    2014\,Apr\,$+1.6$ & 94 & 20 & 1264 & 27467 \\
    2014\,Apr\,$-26$ & 94 & 20 & 2490 & 57154\\
    2014\,Apr\,$-55$ & 96 & 20 & 550 & 13858 \\
    2014\,Jul\,$+1.6$ & 96 & 20 & 785 & 17046 \\
    2014\,Jul\,$-26$ & 96 & 20 & 2062 & 47742 \\
    2014\,Jul\,$-55$ & 96 & 20 & 1806 & 49244 \\
    2014\,Aug\,$+1.6$ & 96 & 20 & 1343 & 29588 \\
    2014\,Aug\,$-26$ & 96 & 20 & 3203 & 75593 \\
    2014\,Aug\,$-55$ & 96 & 20 & 3466 & 98358 \\
    2014\,Oct\,$+1.6$ & 96 & 20 & 943 & 20886 \\
    2014\,Oct\,$-26$ & 96 & 20 & 1719 & 40221 \\
    2014\,Oct\,$-55$ & 96 & 20 & 2049 & 57397 \\ \hline
  \end{tabular}
  \label{tab:sourcenumbers}
  \end{minipage}
\end{table*}

\subsection{Measuring Position Offsets}\label{sec:methodpos}
\begin{figure}
  \centering
  \includegraphics[width=0.48\textwidth]{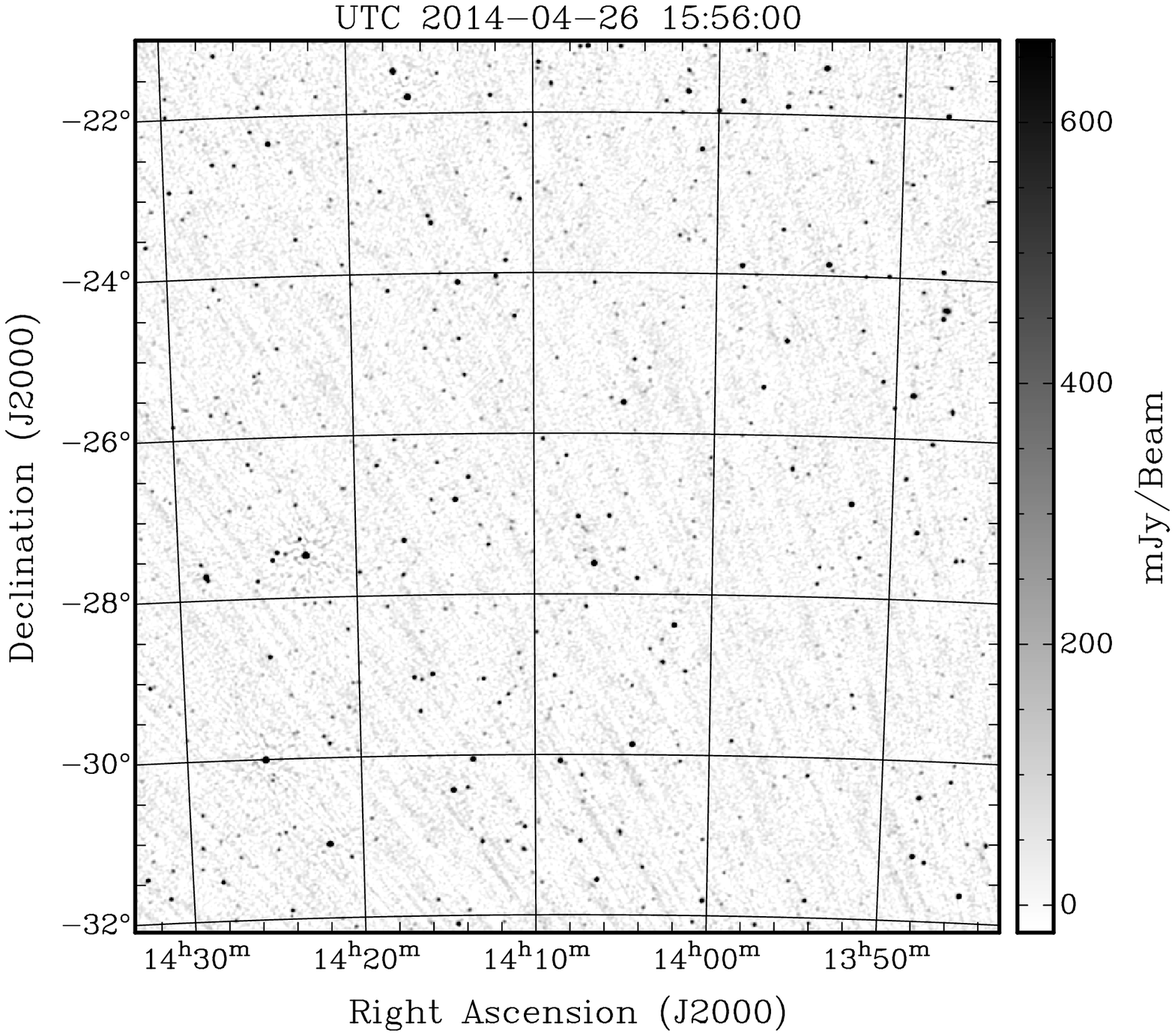}
  \includegraphics[width=0.5\textwidth]{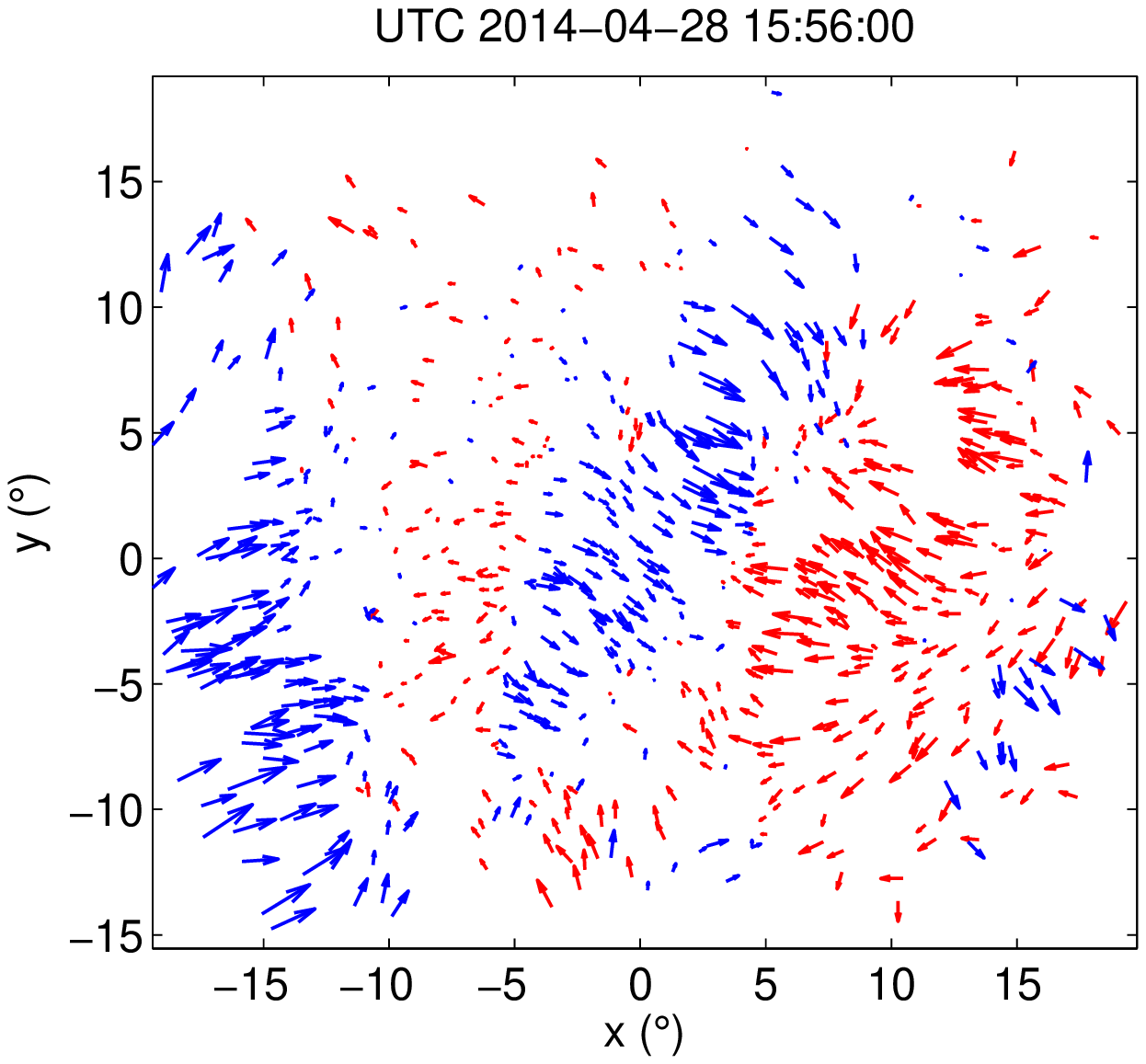}
  \caption{Top panel: An example 2-min snapshot image from 2014\,Apr\,$-26$, zoomed in to the central quarter so that individual sources are visible. The full image contains of order 10$^3$ sources above 5$\sigma$. Bottom panel: The vector field of angular displacements for that snapshot, shown over the whole FoV for sources appearing in at least 20 snapshots. The middle of each arrow marks the location of a point source, and the arrow represents the displacement from its time-averaged position, with arrow lengths scaled to 150 times the actual displacement distance. The $x$-axis points west, the $y$-axis points north, and $(0,0)$ marks the location of the zenith \citep[a Sanson-Flamsteed projection scheme has been used; for details, see][]{Loi2015a_mn2e}. Blue and red denote arrows with positive and negative $x$-components, respectively, to aid visualisation. Organised patches of motion are clearly evident, corresponding to density structures with length scales of $\sim$50\,km for an assumed altitude of 300\,km.}
  \label{fig:arrowplot}
\end{figure}

We measured position offsets by computing the angular displacement of each source (uniquely identified by its NVSS or SUMSS catalogue name) from its time-averaged position in a given dataset. This yielded a set of displacement vectors as a function of time for each source. Use of the time-averaged position as a reference subtracts away fluctuations that are static with respect to the celestial sky. Ionospheric fluctuations would tend to be fixed to the terrestrial sky, while fluctuations fixed to the celestial sky are more likely to be imaging artefacts (e.g.~caused by sidelobes of bright sources). In the absence of other causes of time-dependent fluctuation, any statistically significant scatter in the measured positions occurring within the time span of an observation can be attributed to the ionosphere (tropospheric delays are only $\sim$1\% those of the ionosphere and can be neglected). A single snapshot image from one of the MWATS datasets, along with the associated distribution of angular offset vectors, is shown in Fig.~\ref{fig:arrowplot} (further inspection of the temporal behaviour reveals that a small-scale TID is passing overhead). Organised density fluctuations are widely observed in the remaining datasets, with TIDs and field-aligned ducts being among those visually identifiable \citep[cf.][]{Loi2015a_mn2e, Loi2015_mn2e}. However, it is not our intention to pursue the physics of the underlying phenomena here, but to focus on the broad statistical properties of the fluctuations that are relevant to astrophysical observables.

The largest-scale TIDs, with periods of $\sim$1\,hr \citep{Hunsucker1982}, have wavelengths larger than the MWA FoV and would be removed by conventional calibration approaches (these are only effective at removing ionospheric fluctuations larger than the FoV). The greatest contributors to any residual position offsets in MWA data (following standard calibration approaches) would be those with spatial scales between 10--100\,km. These include small- to medium-scale TIDs, and whistler ducts. The associated timescales of position fluctuation due to these irregularities are of the order minutes to several tens of minutes, significantly shorter than the restriction on the observing duration of a source for analysis (20 snapshots = 2\,hr). Measuring offsets with respect to the time-averaged position thereby isolates these short-term fluctuations. A second reason for choosing to use the time-averaged position rather than the catalogue position as the reference is to cancel out the effects of calibration errors in MWATS data, which appear as global offsets \citep[discussed previously by][]{Loi2015a_mn2e}.

\subsection{Measuring Flux Density Variations}\label{sec:methodflux}
\begin{figure}
  \centering
  \includegraphics[width=0.49\textwidth]{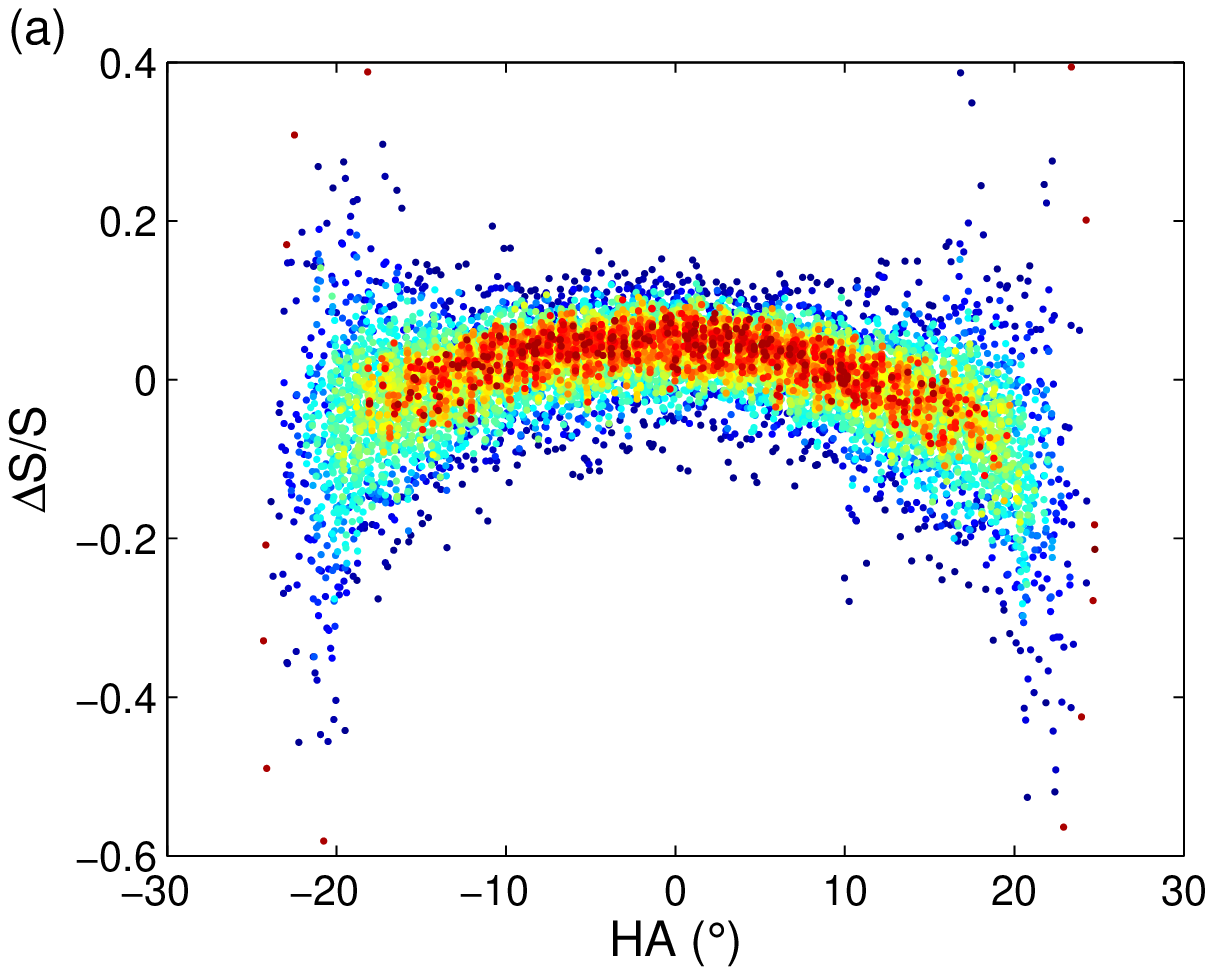}
  \includegraphics[width=0.49\textwidth]{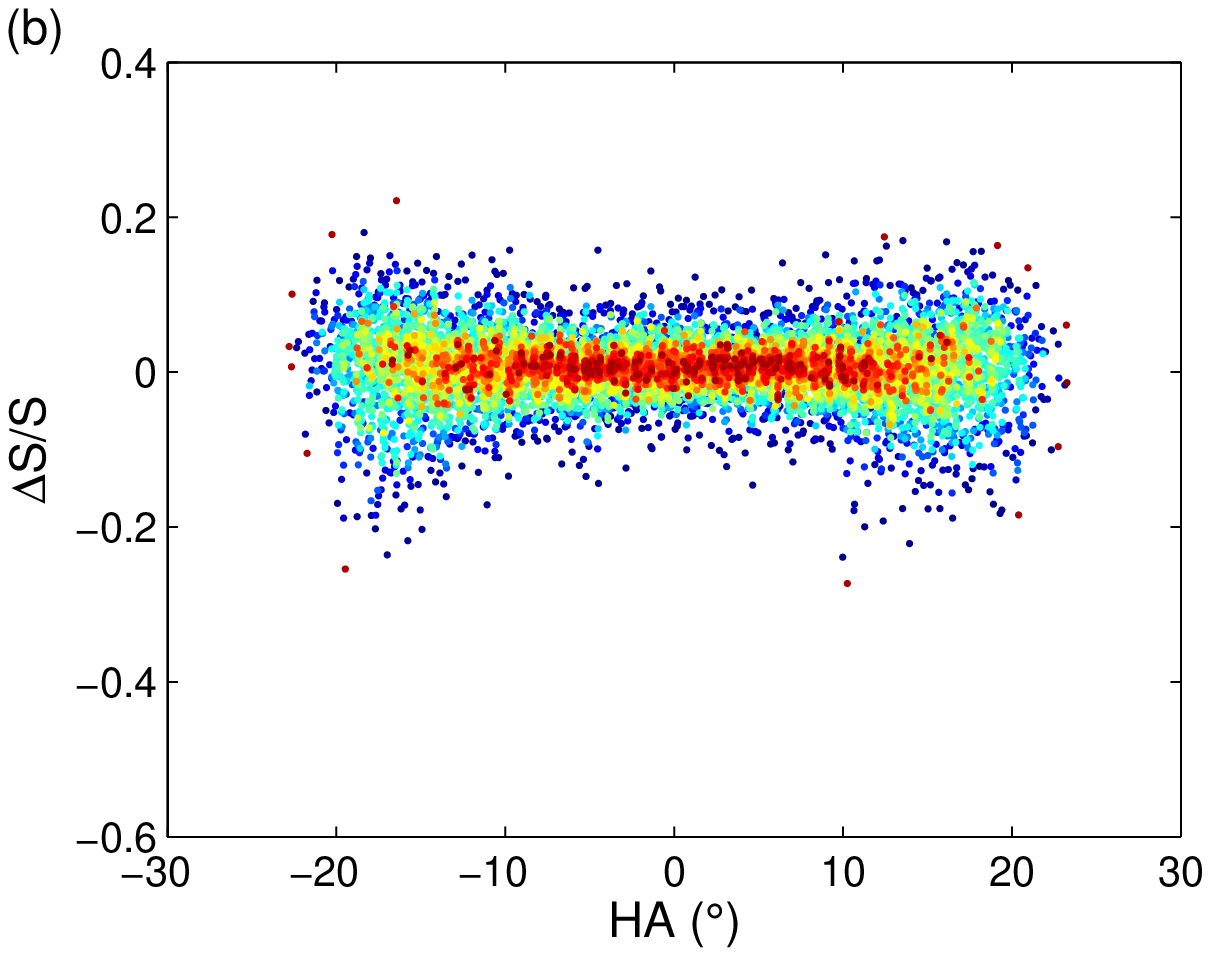}
  \caption{(a) The ensemble statistics of radio light curves (fractional variation from the mean amplitude of the source) from the 2013\,Dec\,$-26$ dataset for all sources brighter than 2\,Jy and appearing in at least 20 snapshots, plotted as a function of hour angle. The pointing is fixed at zenith, and so this equivalently shows the fractional variation in amplitude across the primary beam. Roughly 10\% systematic variations are observed between the centre and FWHM of the beam. Very similar attenuation patterns are seen for all datasets. (b) The resulting light-curve statistics after applying a high-pass filter (removing $\gtrsim$1\,hr trends), plotted on identical axes. The high-pass filtered light curves are those used in subsequent analyses. In both panels, points are coloured by the local density on the page to show the shape of the distribution.}
  \label{fig:lightcurves}
\end{figure}

As argued in \S\ref{sec:overviewflux}, flux density variations associated with ionospheric scintillation are expected to occur on short timescales, generally below the snapshot cadence of MWATS (6\,min). Systematic variations of the peak amplitudes on much longer timescales are likely to be either intrinsic to the source, or a result of instrumental/imaging effects. In all datasets, we detected strong systematic variations in the peak flux densities of unresolved sources as they drifted through the FoV, the fractional attenuation being largest near the edges and smallest near the centre, forming a concave-down modulation pattern that was static with time and present for all datasets. Figure \ref{fig:lightcurves}a illustrates this modulation pattern. This is likely to reflect a residual error in the model of the primary beam used to correct the images. However, we discuss possible explanations involving propagation-related effects later in \S\ref{sec:fluxinterpretation}. 

To remove the concave-down modulation pattern, we applied a high-pass filter to the radio light curves of each individual source. We did this by subtracting a smoothed version of each light curve from itself, where the smoothing interval was set to 9 snapshots (54\,min). The ensemble statistics for raw and high-pass filtered light curves are shown in Fig.~\ref{fig:lightcurves} for all sources seen one of the datasets. The high-pass filtering operation isolates the short-term variations, which we analyse in \S\ref{sec:fluxresults}. Note that Fig.~\ref{fig:lightcurves} does not directly show the flux density $S$ as a function of time, but rather the fractional variation in $S$ with respect to the mean value of each source, as a function of time expressed in terms of the hour angle. Given that the pointing is fixed along the meridian, this equivalently measures E-W position across the FoV. Light curves for different sources are overlaid on top of one another through this transformation.

\section{Results}\label{sec:results}
In this section, we present our findings on the behaviour of angular position offsets and amplitude variations of the sample of sources selected for this study. We did not investigate broadening or shape distortion effects in detail; suffice to say that we did not observe severe distortion or smearing effects on any of the datasets. The shape parameters reported by \textsc{Aegean} for the 2D Gaussian fits to each source were very consistent within and between nights, with axial ratios being 1.1--1.2 on average. We noted the vast majority ($\sim$96\%) of minor axis fits to lie within 10\% of the median value taken over each dataset, with fitting errors likely to account for the few per cent level of scatter. Major axis fits exhibited somewhat greater scatter: $\sim$85\% of fits were within 10\% of the median value for each dataset. This can be explained by the systematic variation of the point-spread function (PSF) major axis over the FoV (caused by the foreshortening of baselines), while the PSF minor axis is expected to be almost constant. The number of sources extracted by \textsc{Aegean} decreased systematically with increasing noise level, consistent with what one might expect from simple considerations (fewer sources can be seen when noise levels are higher). This suggests that broadening/distortion effects due to the ionosphere are insignificant in these datasets and have little effect on the performance of the source finder.

\subsection{Position Offsets}\label{sec:posresults}
\begin{figure}
  \centering
  \includegraphics[width=0.5\textwidth]{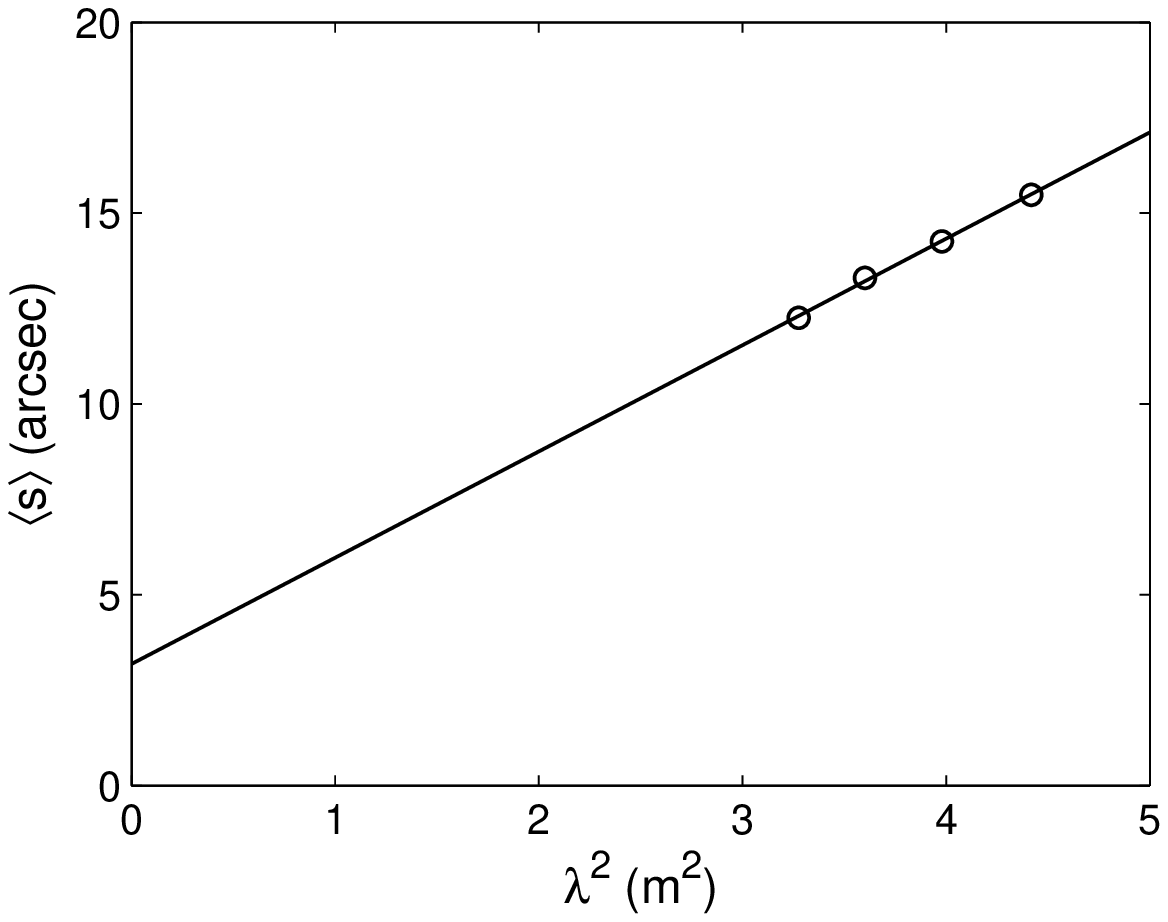}
  \caption{The magnitude of angular position offset of sources in four frequency sub-bands of the 2013\,Dec\,$-26$ dataset, plotted as a function of squared wavelength. Only sources appearing in at least 20 snapshots have been included in the analysis. The line is a linear fit to the four points. The error bars on each data point associated with random fitting errors are smaller than the marker size.}
  \label{fig:offset_vs_lambdasq}
\end{figure}

The angular position offsets display a linear dependence on $\lambda^2$, as shown in Fig.~\ref{fig:offset_vs_lambdasq} for the 2013\,Dec\,$-26$ dataset. Refractive behaviour is expected to follow a $\lambda^2$ proportionality (see Equation \ref{eq:posshift}), but here we see that the intercept is non-zero. This may be explained by a $\sim$3\,arcsec position fitting error, associated with a combination of thermal/confusion noise and inaccuracies in the reference position, that is uncorrelated with the ionospheric errors. In addition, multiple factors with different $\lambda$-dependencies may be contributing to the observed offsets, e.g.~the wavelength dependence of the size of the PSF ($\propto \lambda$) and the sky noise temperature ($\propto \lambda^{2.6}$), which affect the amplitude of scatter in position measurements. However, our data do not have sufficient fractional bandwidth to decompose out possible contributing factors or rule out alternative functional dependences. 

\begin{figure}
  \centering
  \includegraphics[width=0.49\textwidth]{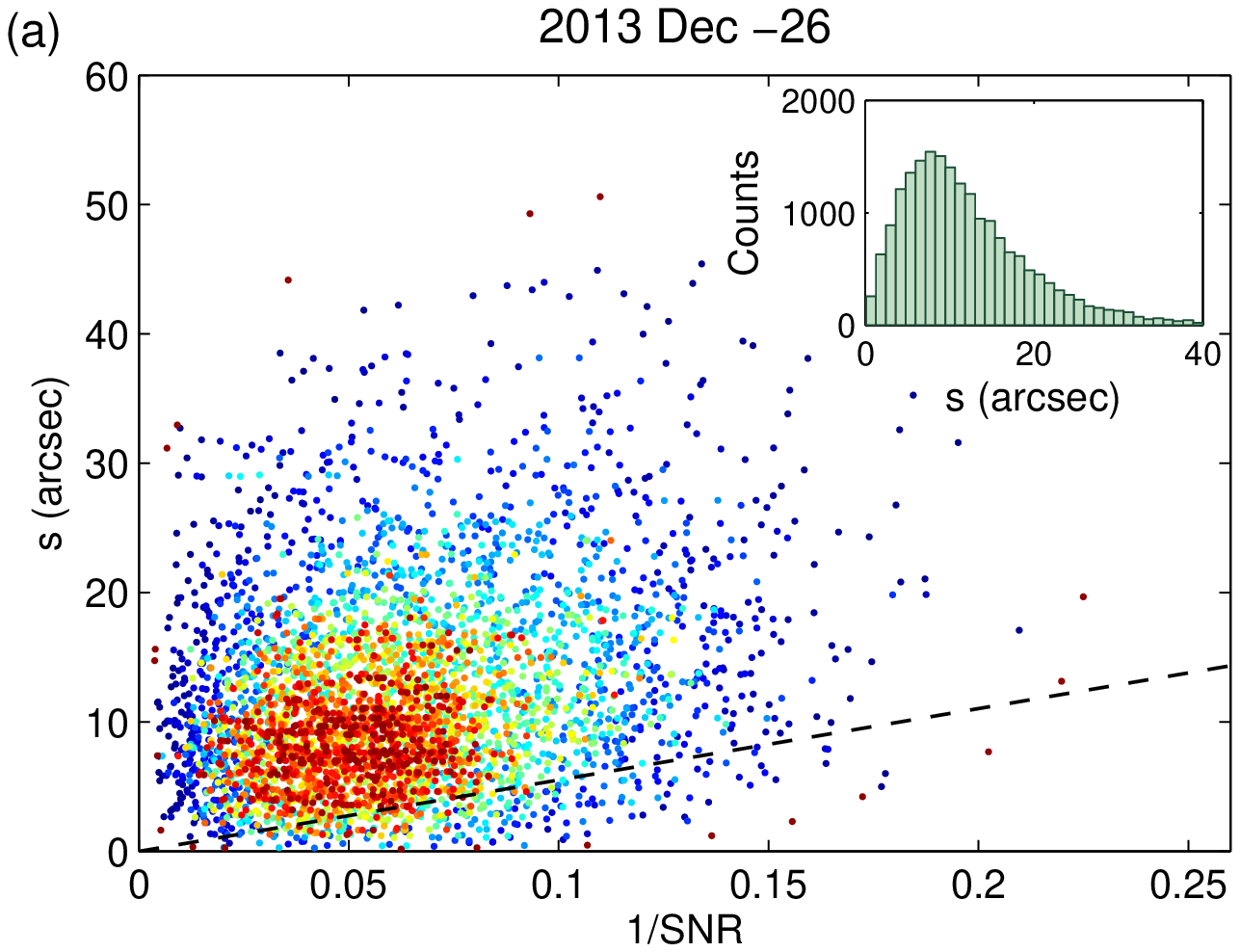}
  \includegraphics[width=0.49\textwidth]{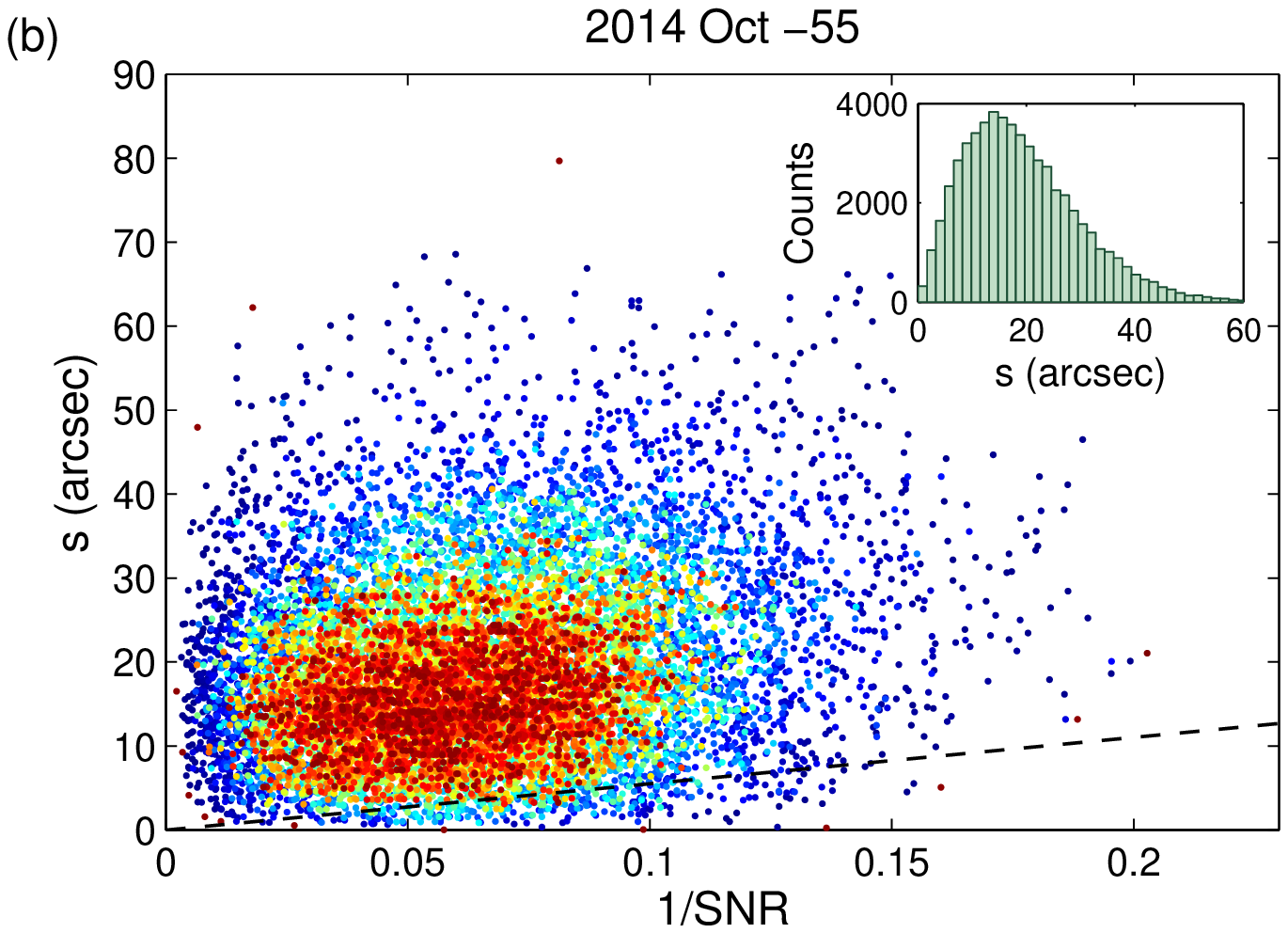}
  \caption{Magnitudes of the position offsets for sources in the (a) 2013\,Dec\,$-26$ and (b) 2014\,Oct\,$+1.6$ datasets, measured with respect to the mean position of the source, against the reciprocal of the S/N ratio. The dotted line indicates the expected offset if these were due to fitting errors. Points are coloured according to the local density on the page to show the shape of the distribution. Only one in five points is plotted. The inset figures show the histograms of the distributions projected onto the vertical axis.}
  \label{fig:offset_vs_SNR}
\end{figure}

Let us denote the angular position offset of a source (measured with respect to the time-averaged position) by the vector $\mathbf{s}$, and let its magnitude be $s \equiv |\mathbf{s}|$. The $s$-values measured for two representative datasets are shown in Fig.~\ref{fig:offset_vs_SNR}. The dashed line marks the expected magnitude of offset if the measured displacements were a consequence of Gaussian fitting errors. We estimated the fitting error for a source of a certain signal-to-noise ratio SNR by \citep{Condon1997}
\begin{align}
  \Delta s \approx \frac{\theta_b}{\mathrm{SNR} \sqrt{8 \ln 2}} \:, \label{eq:poserror}
\end{align}
where $\theta_b$ is the width of the synthesised beam (130\,arcsec at 154\,MHz), and we took SNR to be the ratio between the peak flux density of the source and the local root-mean-square (RMS) noise. 

The observed offsets in all datasets are systematically larger than what one might expect from fitting errors alone. They typically lie in the range 10--20\,arcsec, consistent with the TEC gradients expected for fluctuations driven by atmospheric waves (see \S\ref{sec:overviewposition}). Offsets seldom exceed the pixel size of 45\,arcsec, being sub-pixel around 99\% of the time in most datasets. For the July and August 2014 datasets, higher levels of ionospheric activity have led to a reduction of this rate to 98\% in July and as low as 89\% in August. Mean values of $s$ and the fraction of sub-pixel offsets for each dataset are listed in the second and third columns of Table \ref{tab:results}.

\begin{figure}
  \centering
  \includegraphics[width=0.45\textwidth]{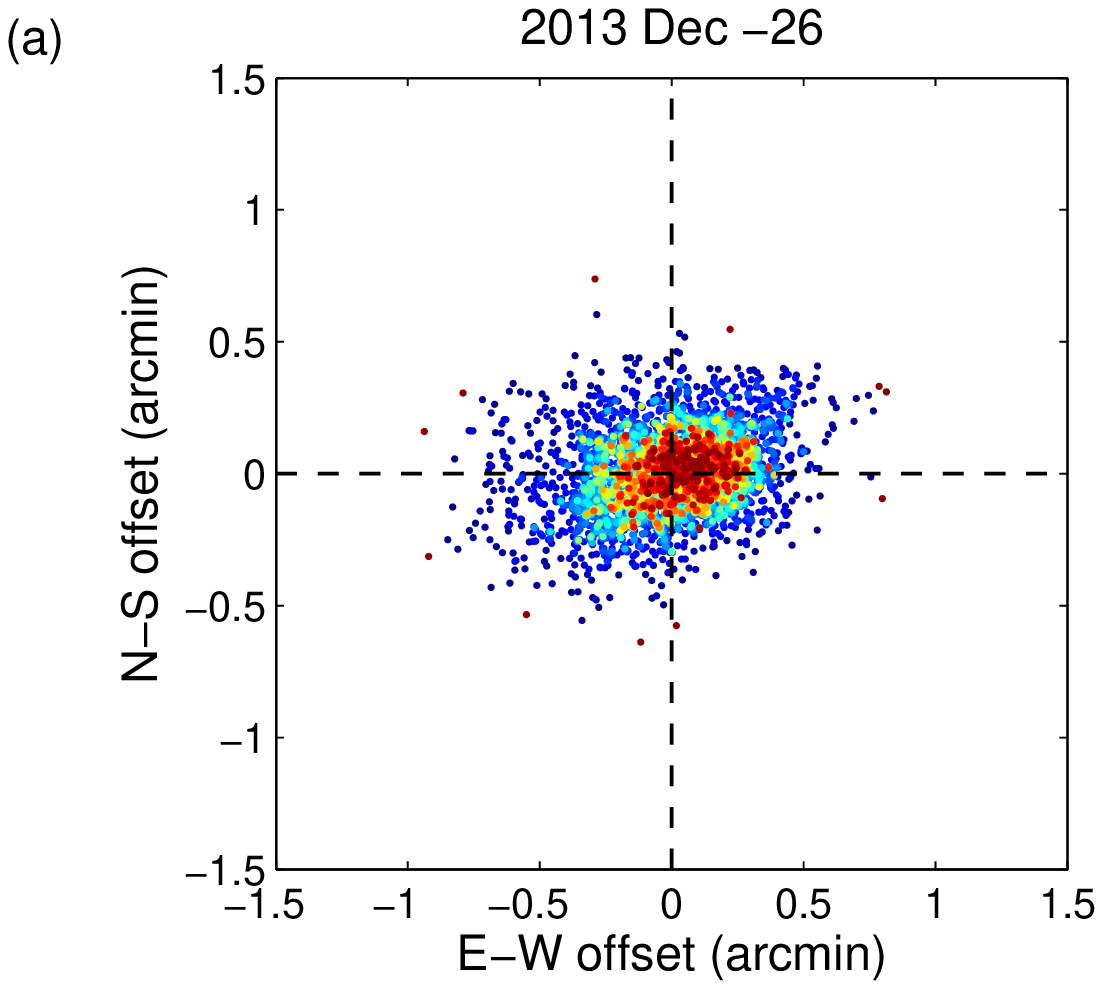}
  \includegraphics[width=0.45\textwidth]{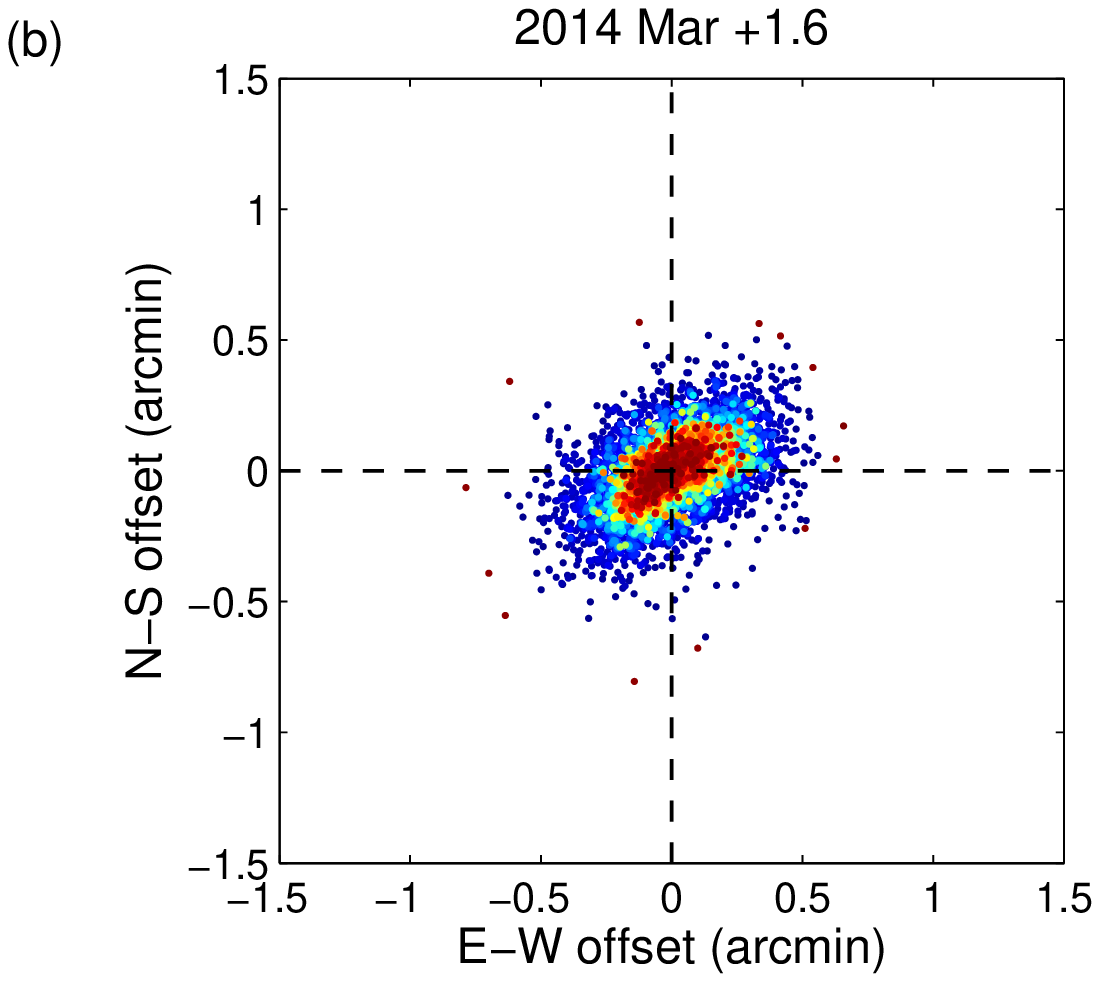}
  \includegraphics[width=0.45\textwidth]{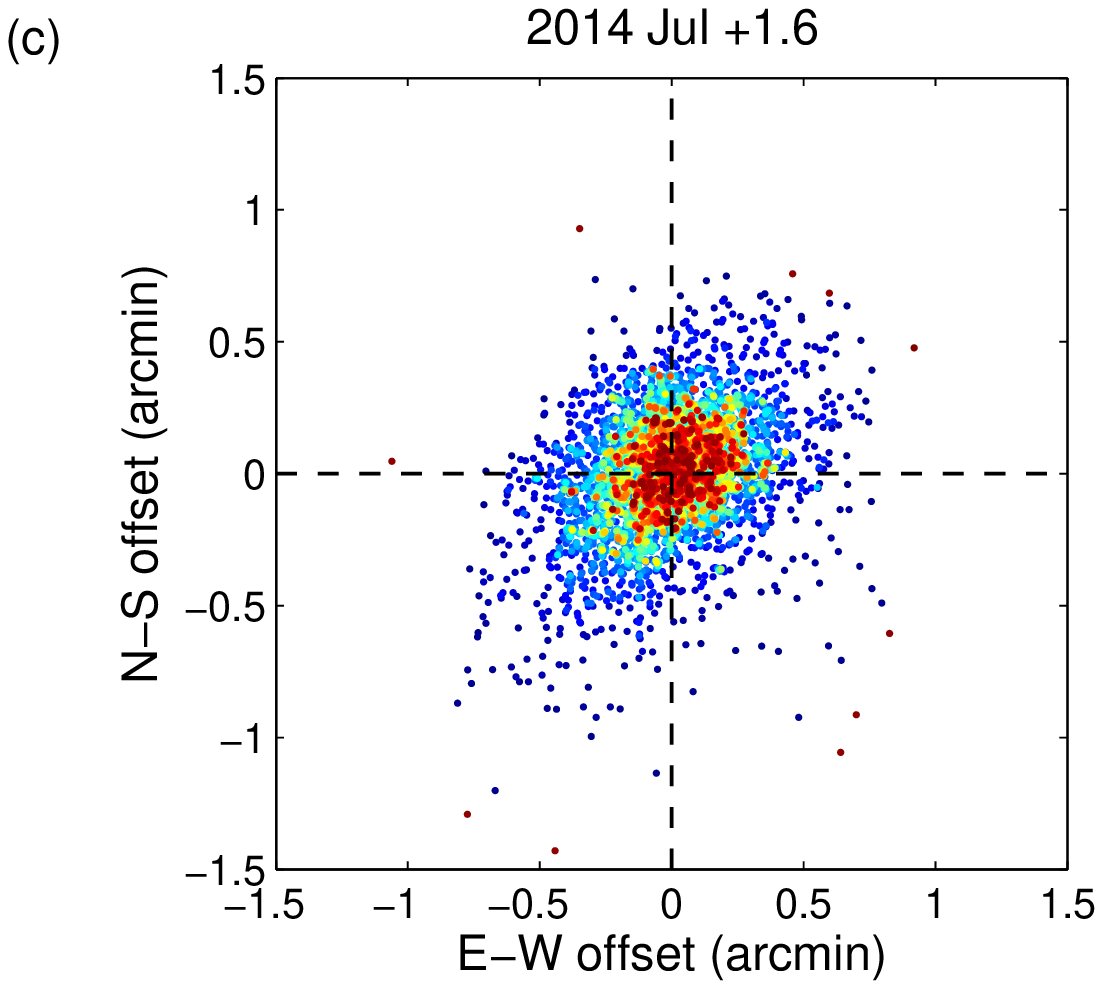}
  \caption{Position offsets for sources in the (a) 2013\,Dec\,$-26$, (b) 2014\,Mar\,$+1.6$ and (c) 2014\,Jul\,$+1.6$ datasets, measured with respect to the time-averaged position of the source. The $x$ axis points west and the $y$ axis points north. The dotted lines mark the location of the origin. Points are coloured according to the local density on the page to show the shape of the distribution. Only one in five points is plotted.}
  \label{fig:offset2D}
\end{figure}

Scatter plots of the position offsets are shown in Fig.~\ref{fig:offset2D}, for three representative datasets. The axes of the plot do not correspond to offsets in RA and Dec but rather offsets along geographic E-W and N-S, computed as described in \citet{Loi2015a_mn2e}. These distributions for the vast majority of datasets show noticeable anisotropy, implying a preferred direction of source displacement. They tend to be elongated into the first and third quadrants, suggesting that fluctuations are preferentially along geographic NW-SE. Elongation of the MWA synthesised beam is insufficient to account for this bias, a point that we discuss in more detail in \S\ref{sec:posinterpretation}. If ionospheric, this indicates that on a long-term basis, there is a statistical preference for TEC gradients to be steeper in the NW-SE direction. However, at any one time, MWA data can exhibit patches of fluctuation oriented in arbitrary directions (e.g.~the waveform shown in Fig.~\ref{fig:arrowplot}b displays some patches of motion directed NE-SW).

\begin{figure}
  \centering
  \includegraphics[width=0.5\textwidth]{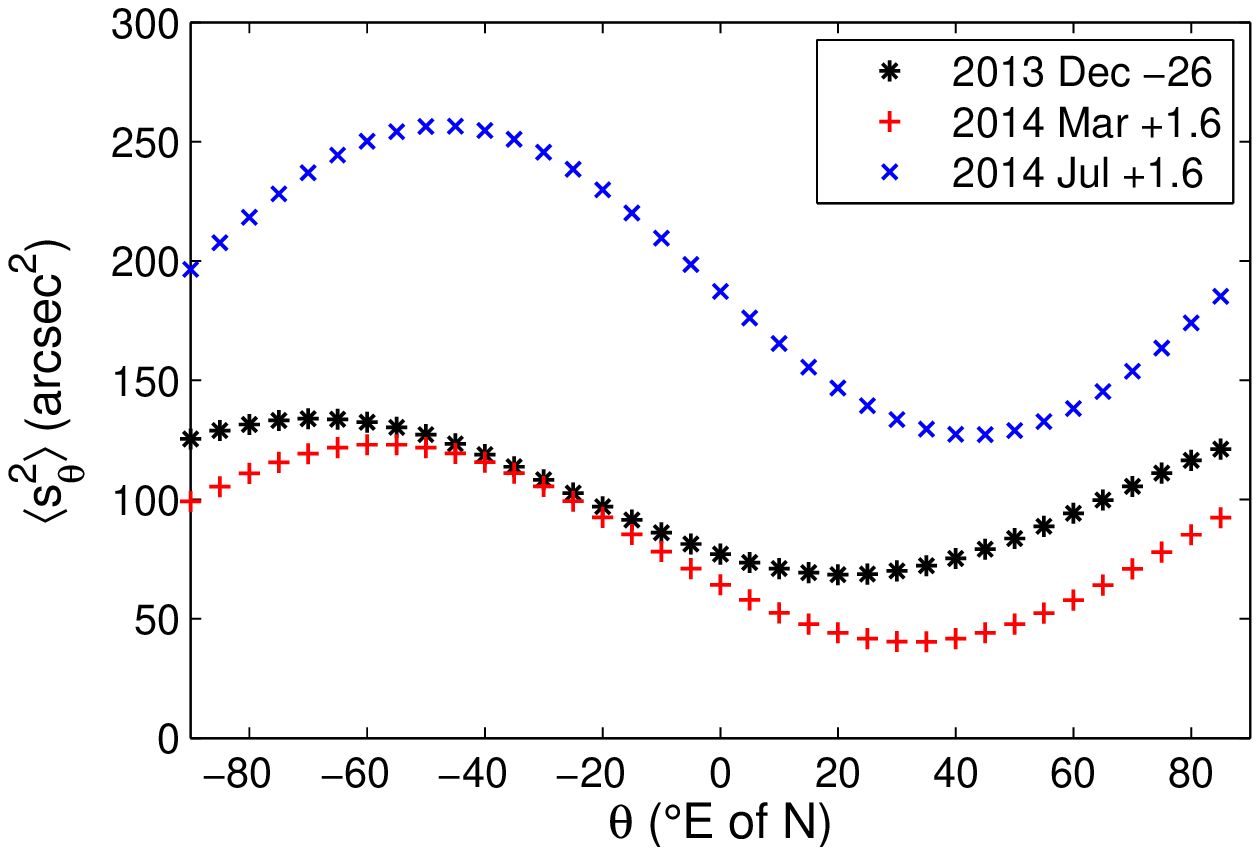}
  \caption{Position fluctuation power (squared projected offset) as a function of the angle of projection $\theta$, measured east of north, for the three datasets shown in Fig.~\ref{fig:offset2D}. The power has been normalised to the number of data points (number of occurrences of any source in any snapshot) in the respective dataset.}
  \label{fig:power_vs_theta}
\end{figure}

The anisotropy in the position offset vectors can be quantified by observing how the scalar projection $s_\theta$ of the offset vectors changes as a function of the projection direction $\theta$. The mathematical details of the steps we took to compute $s_\theta$ are described in \citet{Loi2015a_mn2e}. Briefly, we projected the displacement vectors $\mathbf{s}$ onto a given direction $\theta$ on the sky by taking the inner product with a unit vector pointing in that direction, to obtain a scalar amplitude $s_\theta$. The variation of this amplitude (averaged over all source occurrences) with direction is shown in Fig.~\ref{fig:power_vs_theta} for the three datasets of Fig.~\ref{fig:offset2D}. In analogy with the electric field vectors of light being passed through a polarising filter, the quantity plotted is the average squared amplitude $\langle s_\theta^2 \rangle$ (cf.~intensity). The preferred fluctuation direction is given by the angle $\theta_\mathrm{max}$ at which $\langle s_\theta^2 \rangle$ maximises, and the degree of anisotropy (cf.~linear polarisation) is given by
\begin{align}
  A = \frac{M-m}{M+m} \;,\quad M = \max_\theta \langle s_\theta^2 \rangle \;,\quad m = \min_\theta \langle s_\theta^2 \rangle  \label{eq:anisotropy}
\end{align}
for a given dataset. The values of $A$ and $\theta_\mathrm{max}$ for each dataset are listed in the fourth and fifth columns of Table \ref{tab:results}.

\begin{figure}
  \centering
  \includegraphics[width=0.5\textwidth]{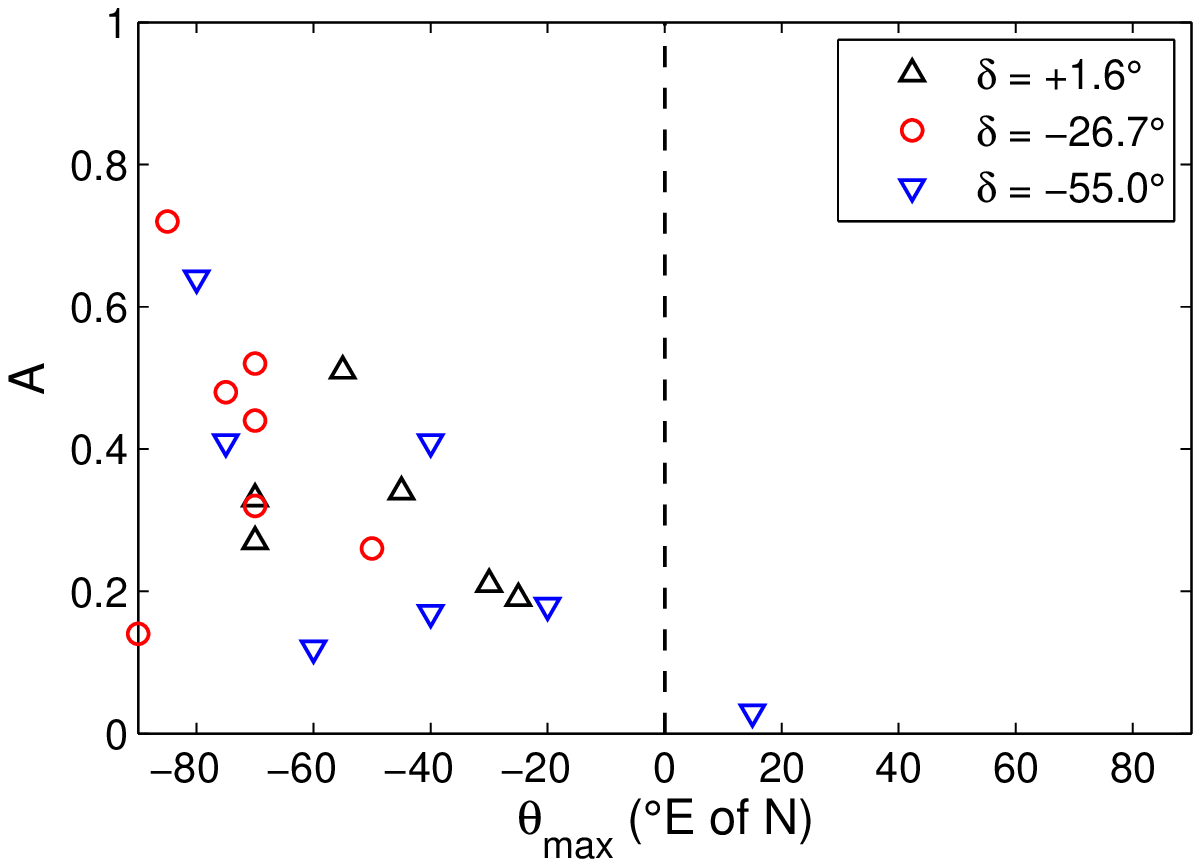}
  \caption{The degree of anisotropy of the position offset vectors, plotted against the direction of greatest position scatter, for each of the 20 datasets. Different symbols have been used to distinguish the different declination bands. The vertical dashed line separates the NW-SE quadrants ($-90^\circ$ to $0^\circ$) from the NE-SW quadrants ($0^\circ$ to $90^\circ$).}
  \label{fig:anisotropy_vs_direction}
\end{figure}

Figure \ref{fig:anisotropy_vs_direction}, which plots the dependence of $A$ on $\theta_\mathrm{max}$, shows that all but one of the 20 datasets have $\theta_\mathrm{max}$ in the NW-SE quadrants. The single dataset with $\theta_\mathrm{max}$ in the NE-SW quadrants has the lowest anisotropy (3\%), meaning that $\theta_\mathrm{max}$ is the least well defined. Discounting this one dataset, there is a clear preference among all the remaining datasets for position offset vectors to be aligned NW-SE ($-90^\circ \leq \theta_\mathrm{max} < 0^\circ$). Furthermore, there appears to be a dependence of $A$ on $\theta_\mathrm{max}$, where for the datasets exhibiting the greatest anisotropy, the direction of the anisotropy is more strongly E-W. The correlation coefficient of the points in Fig.~\ref{fig:anisotropy_vs_direction} is $R^2 = 0.37$, but rises to 0.56 when the outlier at $\theta_\mathrm{max} = -90^\circ$ is excluded.

The characteristic displacement associated with the isotropic (cf.~unpolarised) component of the position offset vector field can be computed as $s_\mathrm{iso} = \sqrt{m}$ (equivalently, the standard deviation of the distribution projected onto its minor axis). If the ionosphere is the only source of systematic anisotropies in the position offset vectors, then $s_\mathrm{iso}$ is an upper bound on the contribution of non-ionospheric sources of astrometric error (e.g.~fitting errors). This is listed for each dataset in the sixth column of Table \ref{tab:results}. Observed values of $s_\mathrm{iso}$ are around 7--10\,arcsec, several times larger than the expected size of fitting errors for typical image sensitivities and source brightnesses ($\sim$3\,arcsec, assuming Gaussian noise statistics).

\subsection{Flux Density Variations}\label{sec:fluxresults}
\begin{figure}
  \centering
  \includegraphics[width=0.5\textwidth]{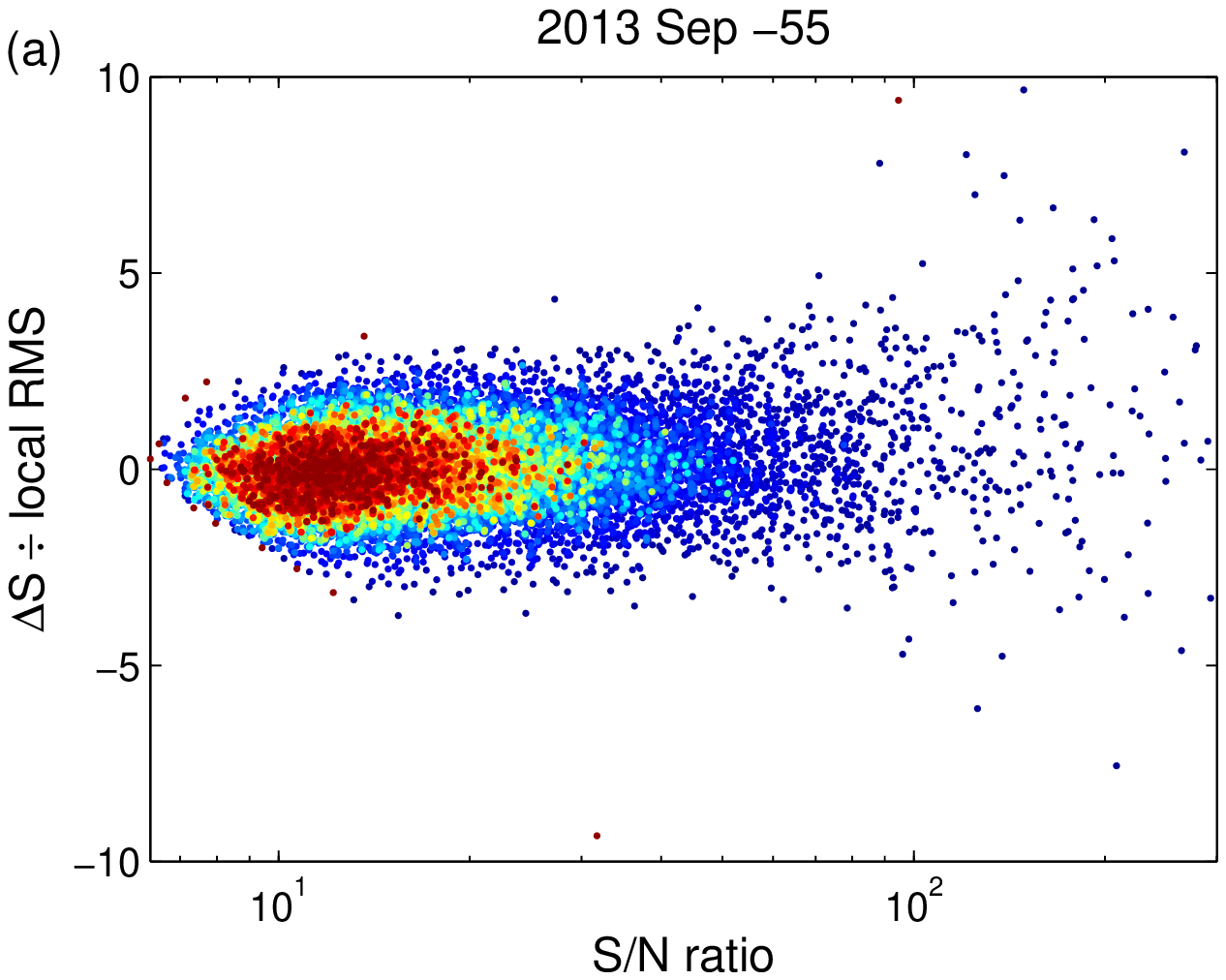}
  \includegraphics[width=0.5\textwidth]{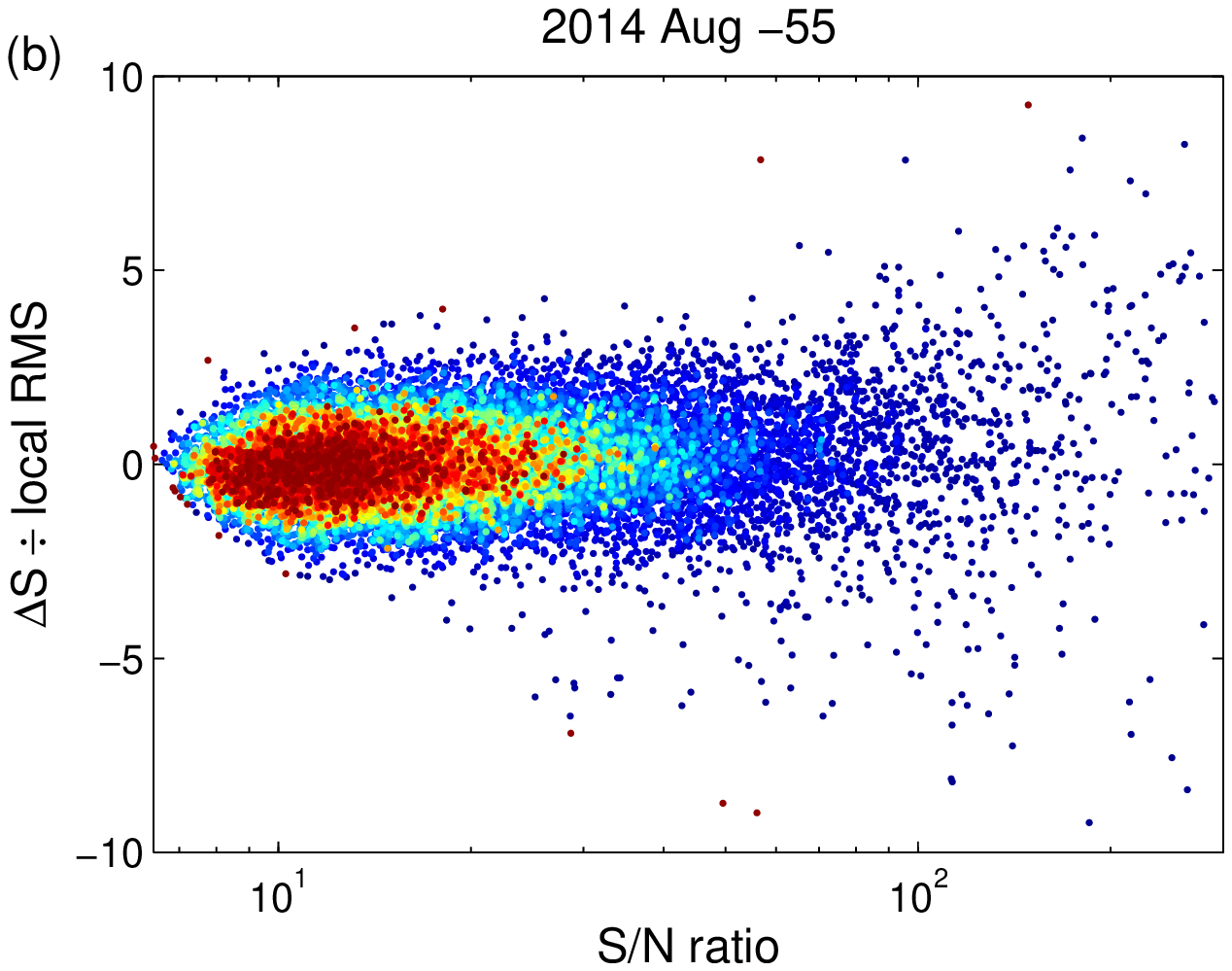}
  \caption{Scatter plots of the difference between measured and mean flux density $\Delta S$ versus the mean flux density $S$ (each axis normalised by the local RMS), for (a) 2013\,Sep\,$-55$ and (b) 2014\,Aug\,$-55$, the two datasets with the lowest and highest average amplitude of position fluctuations, respectively. The horizontal axis is shown on a logarithmic scale. Points are coloured by the local density on the page to show the shape of the distribution. Only one in five points is plotted. Note that the taper (drop in source counts) for S/N ratios below 10, despite faint sources being much more common than bright sources, is due to a sample selection effect: we required that sources appear in a large number of snapshots, biasing our sample to exclude fainter sources that might be lost in regions of high image RMS.}
  \label{fig:fluxdistr}
\end{figure}

For each dataset, we placed an upper limit on the contribution of ionospheric scintillation by analysing the distributions of $\Delta S$ values versus $S$. Here $S$ is the time-averaged flux density of a source and $\Delta S$ is the difference between a flux density measurement and $S$ (after applying a high-pass filter; see \S\ref{sec:methodflux}). Scatter plots of $\Delta S$ versus $S$ for the two datasets exhibiting the highest and lowest amplitudes of position scatter are shown in Fig.~\ref{fig:fluxdistr}, where both quantities are expressed in as multiples of the local root-mean-square (RMS) noise. The RMS values were computed by \textsc{Aegean} based on the interquartile range over an area 20$\times$20 beams in size centred about each source. Each individual $\Delta S$ and RMS value is a function of source and snapshot. 

In the analysis described below, all $\Delta S$ measurements are treated as independent, without consideration of their time ordering. This is in line with our goal here to characterise not celestial but possible terrestrial-based activity occurring on timescales comparable to or shorter than the integration time of each snapshot. We do not perform a light curve-based analysis, which is more relevant to searches for celestial-based variability \citep[e.g.][]{Gaensler2000, Bell2014_mn2e}. It is important to appreciate that the rotation of the Earth causes celestial sight lines to drift with respect to the ionosphere at a significant speed (about 1$^\circ$ between MWATS snapshots at the same Az/El, larger than $r_F$ at 300\,km altitude), and so the radiation from a given source passes through a completely different patch of the ionosphere between adjacent snapshots. We therefore do not expect scintillation to produce substantial temporal correlations in the light curves of individual sources measured at the cadence of MWATS.

Our approach to separating the scintillation component from other contributors to temporal variation relies on the assumption that scintillation effects are described by a characteristic fractional variation in amplitude (modulation index), whereas thermal noise and sidelobe confusion (two important sources of noise in MWA data) are associated with characteristic absolute variations in amplitude. The faintest sources will be dominated by absolute variation effects, while the brightest sources will be dominated by fractional variation effects. Assuming that the different types of error are independent, the vertical spread as a function of signal-to-noise (S/N) ratio is therefore expected to be the quadrature sum of a constant and a linear component.

To decompose these two components, we binned the data by S/N ratio into 19 evenly-spaced intervals up to a maximum S/N ratio of 100 in steps of 5, with the exception of the first bin which included all points with S/N ratio less than 10. We computed the variance $V$ of $\Delta S/\mathrm{RMS}$ values in each bin and then performed a least-squares fit to $V = V(X)$, where $X$ was taken to be the mean S/N ratio of points in a bin. We assumed the functional form $V(X) = \kappa^2 + \mu^2 X^2$, with $\kappa^2$ and $\mu^2$ unconstrained parameters. Absolute-fluctuation effects are quantified by $\kappa$ (expressed as a multiple of the local RMS), and fractional-variation effects by $\mu$ (which can be interpreted as a modulation index). In all datasets we obtained positive values for $\kappa^2$ and $\mu^2$, implying real values for $\kappa$ and $\mu$ and non-zero amounts of each type of noise. Values of $\kappa$ ranged between 0.7 and 0.9, consistent with there being a significant level of classical confusion in the images (much of the pixel-to-pixel RMS is static), while values of $\mu$ ranged between 1--3\% (these are listed in Table \ref{tab:results} for each dataset). The values of $\mu$ obtained are an upper bound on the modulation index associated with scintillation. We discuss the implications of this in \S\ref{sec:fluxinterpretation}. We did not find any strong correlations between $\mu$ and $\langle s \rangle$, the median RMS, the median S/N ratio or $\theta_\mathrm{max}$.

For this analysis we have chosen to ignore the static, concave-down modulation of the peak flux density (e.g.~Fig.~\ref{fig:lightcurves}a), removing this component from the light curves using a high-pass filter. We do not believe that this is due to the ionosphere, but rather that it is the residual of an imperfect primary beam model. Our quantitative arguments against this being ionospheric in origin are presented in \S\ref{sec:fluxinterpretation}.

\begin{table*}
  \centering
  \begin{minipage}{15cm}
  \caption{Results for the 20 MWATS datasets. The second to last columns, from left to right, correspond to the mean displacement magnitude $\langle s \rangle$ averaged over all sources in the dataset, the percentage $f_\mathrm{subpix}$ of displacements that are sub-pixel (pixel size 45\,arcsec), the degree of anisotropy $A$, the direction of greatest position scatter $\theta_\mathrm{max}$ (measured east of north), the amplitude of displacement $s_\mathrm{iso}$ associated with the isotropic component of the position offset vector field, the upper bound $\mu$ on the modulation index associated with scintillation, and the characteristic position angle of the PSF (east of north).}
  \begin{tabular}{lS[table-format=3.2]S[table-format=3.2]rrrrrr} \hline
    Dataset & \multicolumn{1}{c}{$\langle s \rangle$ (arcsec)} & $f_\mathrm{subpix} \SI{}{(\percent)}$ & $A$ & $\theta_\mathrm{max}$ ($^\circ$) & $s_\mathrm{iso}$ (arcsec) & $\mu$ (\%) & bpa ($^\circ$) \\ \hline
    2013\,Sep\,$-55$ & 9.8 & 99.98 & 0.18 & $-20$ & 7.3 & 1.9 & 81 \\
    2013\,Oct\,$-26$ & 13.5 & 99.03 & 0.44 & $-70$ & 8.7 & 2.1 & 160 \\
    2013\,Dec\,$+1.6$ & 12.8 & 99.44 & 0.27 & $-70$ & 9.2 & 1.5 & 113 \\
    2013\,Dec\,$-26$ & 12.0 & 99.79 & 0.32 & $-70$ & 8.3 & 1.5 & 38 \\
    2013\,Dec\,$-55$ & 13.6 & 99.75 & 0.12 & $-60$ & 10.4 & 1.4 & 85 \\
    2014\,Mar\,$+1.6$ & 10.8 & 99.90 & 0.51 & $-55$ & 6.4 & 1.7 & 41 \\
    2014\,Mar\,$-26$ & 11.5 & 99.95 & 0.52 & $-70$ & 6.6 & 1.7 & 20 \\
    2014\,Mar\,$-55$ & 13.6 & 99.84 & 0.03 & $+15$ & 10.8 & 2.6 & 85 \\
    2014\,Apr\,$+1.6$ & 10.3 & 99.80 & 0.19 & $-25$ & 8.0 & 1.5 & 23 \\
    2014\,Apr\,$-26$ & 9.9 & 99.89 & 0.26 & $-50$ & 7.3 & 1.4 & 20 \\ 
    2014\,Apr\,$-55$ & 11.8 & 99.76 & 0.41 & $-40$ & 7.6 & 1.9 & 59 \\ 
    2014\,Jul\,$+1.6$ & 16.4 & 97.86 & 0.34 & $-45$ & 11.3 & 3.4 & 149 \\
    2014\,Jul\,$-26$ & 15.2 & 97.86 & 0.48 & $-75$ & 9.4 & 2.4 & 20 \\
    2014\,Jul\,$-55$ & 16.6 & 98.15 & 0.41 & $-75$ & 10.5 & 2.3 & 38 \\
    2014\,Aug\,$+1.6$ & 14.7 & 97.90 & 0.33 & $-70$ & 10.6 & 3.0 & 160 \\
    2014\,Aug\,$-26$ & 17.2 & 94.21 & 0.72 & $-85$ & 8.8 & 2.0 & 5 \\
    2014\,Aug\,$-55$ & 22.8 & 89.4 & 0.64 & $-80$ & 12.5 & 2.1 & 45 \\
    2014\,Oct\,$+1.6$ & 11.5 & 99.86 & 0.21 & $-30$ & 8.6 & 1.9 & 124 \\
    2014\,Oct\,$-26$ & 13.5 & 99.62 & 0.14 & $-90$ & 10.4 & 2.0 & 142 \\
    2014\,Oct\,$-55$ & 19.3 & 97.72 & 0.17 & $-40$ & 14.2 & 2.9 & 117 \\ \hline
  \end{tabular}\\
  \label{tab:results}
  \end{minipage}
\end{table*}

\section{Discussion}\label{sec:discussion}

\subsection{Interpreting Position Offsets}\label{sec:posinterpretation}
A possible contribution to the anisotropy in the position offset vector field comes from the $u,v$-coverage of the observation, which determines the shape of the MWA PSF characterising the direction-dependent resolution of the instrument. Asymmetries in the PSF can arise from non-uniformities in the $u,v$-coverage. If the $u,v$-coverage produces a PSF that is elongated along a certain direction, then this can increase the scatter in position measurements along that direction independently of ionospheric effects.

We checked to see if the preferred direction of position scatter could be accounted for by the PSF shape and associated fitting errors. We did this by examining the shape parameters fitted to each source by \textsc{Aegean}, and compared the characteristic position angle for the source fits in each dataset with the direction of greatest position scatter. The restoring beam is forced to be a circular Gaussian, and so the restoration step should introduce no additional anisotropy in the shapes of sources. However, sources in general will not be completely deconvolved and so residual components of the flux density will cause sources to be slightly anisotropic, reflecting the shape of the underlying PSF. For most datasets we observed the distribution of position angles to peak at a certain value, which we took to be the characteristic PSF position angle for that dataset. These values are listed in the last column of Table \ref{tab:results}.

\begin{figure}
  \centering
  \includegraphics[width=0.5\textwidth]{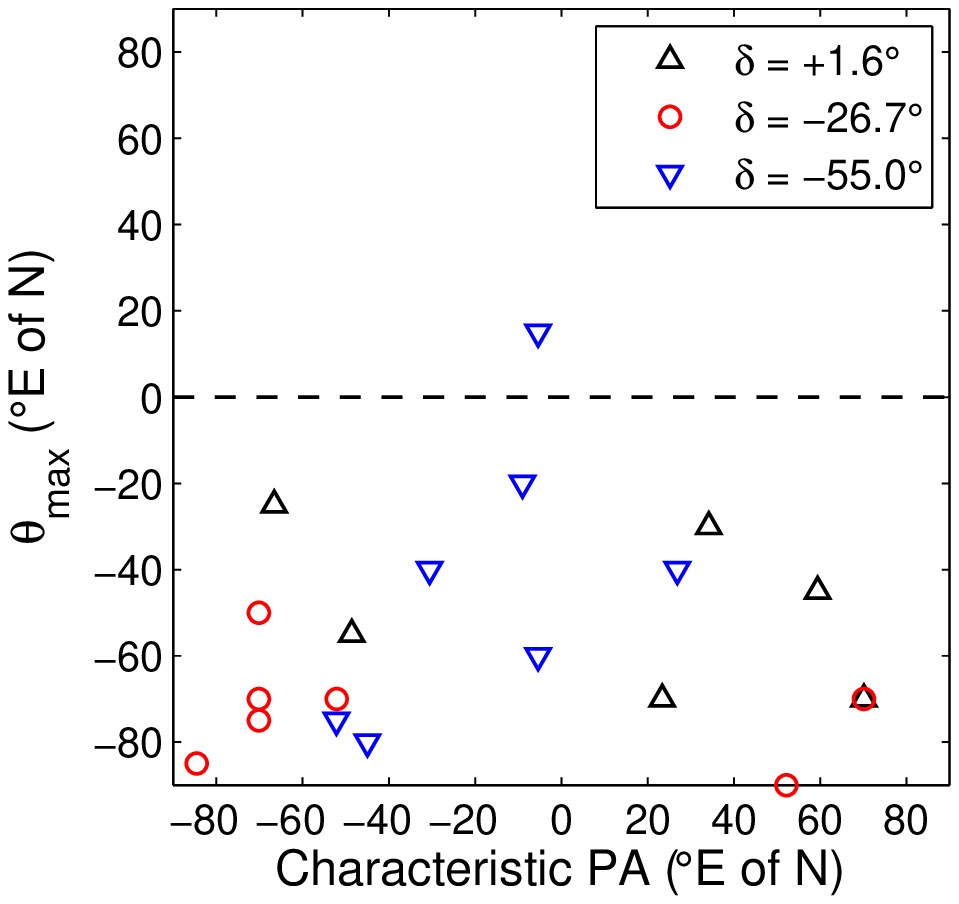}
  \caption{The direction of greatest angular position scatter, plotted against the characteristic position angle of Gaussian fits to sources in the dataset, for each of the 20 datasets. Different symbols have been used to distinguish the three declination bands. Recall that $\delta = -26\fdg7$ corresponds to a zenith pointing, while $\delta = +1\fdg6, -55\fdg0$ are pointings $\sim$30$^\circ$ off zenith towards the north and south. A horizontal line has been drawn at 90$^\circ$ separating the NW-SE quadrants ($-90^\circ$ to $0^\circ$) from the NE-SW quadrants ($0^\circ$ to $90^\circ$).}
  \label{fig:scatter_vs_pa}
\end{figure}

Figure \ref{fig:scatter_vs_pa} plots the direction of greatest scatter against the characteristic position angle of source shape fits. It can be seen that position angles differ substantially between datasets, spanning the full range of values from $-90^\circ$ to $+90^\circ$. They are not correlated with the directions of greatest scatter, which are largely confined between $-90^\circ$ and $0^\circ$. Gaussian fitting errors arising from an asymmetric PSF therefore cannot account for the preferential NW-SE offsets, which require a separate explanation.

The observation that the PSF position angles for off-zenith pointings are closer to $0^\circ$ (N-S) than for zenith pointings is consistent with the foreshortening of the projected baselines when the telescope is pointed towards the north or south. This shrinks the $u,v$-coverage and elongates the PSF in the N-S direction. In the case of the zenith pointing, it can be seen that position angles tend to be closer to E-W. This can be explained by the fact that the array is slightly more extended in the N-S than E-W direction \citep{Tingay2013_mn2e}. Night-to-night variations in the PSF shape are determined by which tiles are flagged/unavailable at the time of observation, and also ionospheric conditions. This may account for the spread in characteristic position angles between observations at the same Az/El.

The tendency for position offsets to be preferentially NW-SE is consistent with reports of similar anisotropies in the literature, as detected in airglow measurements of mid-latitude ionospheric density structures \citep{Martinis2006, Pimenta2008}. It has been noted that density structures in the southern hemisphere tend to be preferentially elongated NE-SW, giving rise to TEC gradients steepest along NW-SE. This agrees with the statistical preference for NW-SE offsets seen in our MWA data. It is speculated that an electrodynamic instability known as the Perkins instability \citep{Perkins1973, Miller1997, Hamza1999, Yokoyama2009} may account for the formation of structures elongated preferentially along a certain direction. Under this mechanism, vertical undulations (e.g.~seeded by atmospheric waves) alter the Pedersen conductivity and supporting forces in a self-reinforcing manner, causing an initial perturbation to be amplified. The growth rate of this instability is direction-dependent with respect to the ambient electric and magnetic fields. However, an investigation into whether or not the perturbations detected by the MWA indeed form as a result of the Perkins instability is beyond the scope of this work. 

That the position offsets match the known preferred direction of ionospheric structures at mid-latitudes, but show no correlation with the PSF position angle, supports the idea that a significant amount of the scatter in angular position measurements is indeed due to ionospheric refraction. A further 20--30\% of the overall scatter may be explained by Gaussian fitting errors.

\subsection{Interpreting Flux Density Variations}\label{sec:fluxinterpretation}
In quantifying extrinsic amplitude variability, we have relied on the assumption that the sources used in this study are intrinsically non-variable. While this is true of the vast majority of celestial radio sources, it is of course possible for some to produce transient and/or variable radio emission. Transient sources, specifically those appearing for less than an hour, would have been excluded by our decision to analyse only sources appearing at least 10 or 20 snapshots. Individual variable sources may have been present, but these are neither the focus of detection in this study nor likely to affect the measured bulk trends of the population, and so we have ignored their possible existence.

Physical conditions for the formation of scintillation-inducing irregularities are likely to occur over regional-scale patches, and so scintillation events are expected to affect large numbers of celestial sources in the FoV at once. This is an assumption behind our approach to quantifying ionospheric effects by examining bulk trends in variability. However, we cannot rule out the possibility that scintillation-causing irregularities form in localised ($\sim$1\,km-wide) patches of the ionosphere, thereby only affecting isolated sources at any one time. We have not inspected our data for evidence of this. In any case, without extensive follow-up observations, it would be difficult to distinguish such cases from intrinsic variability. The search for intrinsic variability in MWATS is the subject of separate ongoing work.

The finding that $\mu$ lies between 1--3\% indicates that besides thermal noise and sidelobe confusion, there is an additional contribution to the amplitude variation of each source of order several per cent the source brightness. Some of this could be ionospheric scintillation, but gain calibration errors and residual primary beam attenuation may also enter as fractional variation effects. The $\mu$-values are therefore an upper bound on, rather than an unbiased estimate of, the contribution from scintillation. Decomposing these down into the various possible contributors is a task we defer to future work. For a 1-Jy source, $\mu \sim$~1--3\% corresponds to a 10--30\,mJy additional fluctuation, which is comparable to the typical image RMS in MWATS. Most sources are fainter than this in which case thermal noise and sidelobe confusion effects will be more important, but for brighter sources this contribution will be noticeable and may need to be appropriately incorporated into analyses of light-curve variability.

The bound on $\mu$ enables us to constrain the diffractive scale $r_\mathrm{diff}$, which can be thought of as the characteristic length scale of ionospheric phase fluctuations (physical distance within which phases vary by 1\,rad). We apply a number of simplifying assumptions: (1) that the MWA is compact enough to assume intensity (i.e.~zero-baseline) scintillations\footnote{We justify this by the fact that the MWA is centrally condensed: 112 out of 128 tiles lie within a 750\,m radius of the core, and so most baselines and baseline separations are shorter than $r_F$.}, and (2) that the spectrum of phase fluctuations is described by the Kolmogorov power law \citep{Kolmogorov1941}. A modulation index of much less than unity indicates weak scintillation, in which case we have the result that $\mu \approx (r_F/r_\mathrm{diff})^{5/6}$ \citep{Narayan1992}. As argued in \S\ref{sec:overviewflux}, the decorrelation timescale for weak ionospheric scintillation is upwards of 10\,s. The integration time for MWATS is thus $\sim$10 decorrelation timescales, suppressing $\mu$ by a factor of up to 3--4. Substituting the values for $r_F \sim 1$\,km and the observed $\mu \sim$ 1--3\%, we obtain $r_\mathrm{diff} \gtrsim$ 20\,km. This is consistent with being in the weak scintillation regime ($r_F < r_\mathrm{diff}$), and also matches the sizes of structures detected in MWA data \citep{Loi2015a_mn2e, Loi2015_mn2e}. This suggests that the empirical value of $\mu$ is physically reasonable.

\begin{figure}
  \centering
  \includegraphics[width=0.5\textwidth]{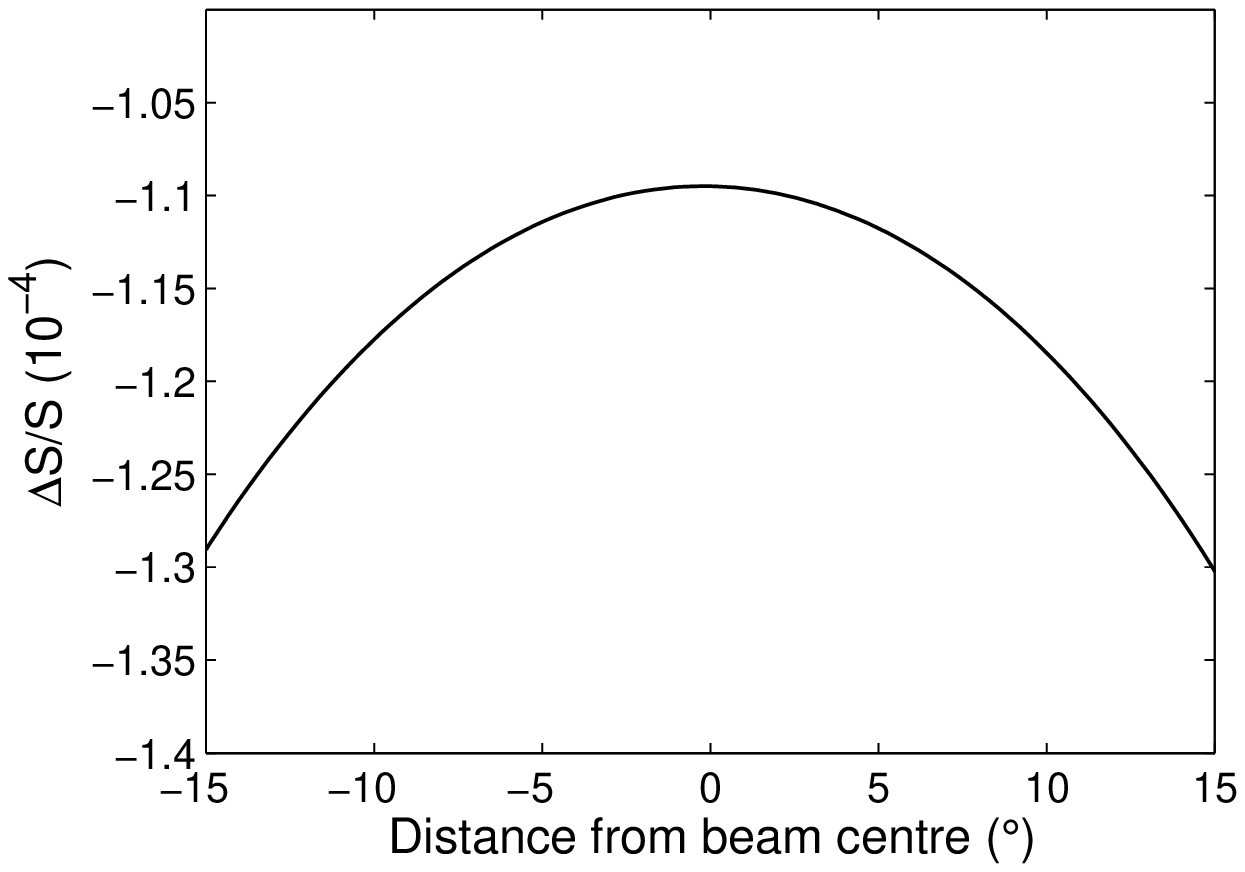}
  \caption{The predicted fractional deviation in peak flux density (compared to the true peak flux density) of a source as a function of position across the FoV for a zenith drift scan, due to a smooth ionosphere located at a constant altitude of 300\,km. This calculation assumed an integration time of 112\,s, an observing frequency of 154\,MHz, a background TEC of 10\,TECU and a PSF width of 130\,arcsec, with the source passing directly overhead. Note that the vertical axis is given in multiples of $10^{-4}$.}
  \label{fig:seczeta_smearing}
\end{figure}

Systematic modulations in the amplitudes of celestial sources can be produced by propagation effects, but we now demonstrate that these are not sufficiently large to explain the observed concave-down pattern in the light curves (Fig.~\ref{fig:lightcurves}a). A still, uniform, constant-altitude ionosphere will induce smearing of radio sources as they drift overhead, because the path length through the ionosphere varies as a function of viewing direction. This variation is non-linear, implying that $\nabla_\perp \mathrm{TEC}$ varies across the FoV. The resulting smearing (since the offset is proportional to $\nabla_\perp \mathrm{TEC}$) lowers the peak flux density relative to its true value. The estimated magnitude of this effect for MWATS observing parameters is plotted in Fig.~\ref{fig:seczeta_smearing}. Although the shape qualitatively matches the observed concave-down pattern, the amplitude is about three orders of magnitude too small and this effect therefore cannot account for observations.

Consider instead the possibility that the attenuation is a result of a propagation effect, such as scattering or absorption, that has a constant probability per unit length of interacting with the incoming signal. Let $f(\zeta)$ denote the strength of the transmitted signal relative to the original signal for a path at zenith angle $\zeta$. The light curves indicate that $f(15^\circ)/f(0^\circ) \approx 0.9$, and given that $f(\zeta) = f(0)^{\sec \zeta}$ we can then solve for $f(0) = 0.05$. This means that for the attenuation pattern to be explained by a constant-probability-per-unit-length scattering/absorption effect, the decrease in signal amplitude needs to be by a factor of about 20 at zenith. This is significantly larger than can be attributed to absorption, which is $\lesssim$1\% at 100\,MHz \citep{Thompson2001}. If it were caused by scattering due to small-scale irregularities, then by conservation of total (integrated) flux density this should produce highly scatter-broadened sources (larger by a factor of 20), which are not observed in the data.

That the concave-down attenuation pattern appears in all datasets and persists over entire nights of observation makes it unlikely that this is due to extreme ionospheric conditions, of the kind that might cause the observed distortion via the above effects. It is more likely that these systematic variations are instrumental in nature, arising from imperfect knowledge of the MWA primary beam.

\subsection{Practical Implications}
We have established that the typical scatter in angular position is 10--20\,arcsec at 154\,MHz. Although dominated by ionospheric refraction and several times larger than estimated fitting errors, this is sub-pixel for MWATS the vast majority ($\gtrsim$99\%) of the time and thereby almost always sub-synthesised beam for the MWA at 154\,MHz. This implies that ionospheric refraction is unlikely to present a major problem for the automated cross-matching of sources for time-domain astrophysics with the MWA at this frequency, as long as a cross-matching radius of $\sim$1--2\,arcmin is used. However, there were several datasets in which higher levels of ionospheric activity causing sources to displace by angular distances of order the synthesised beam or more necessitated a larger cross-matching radius, which we set to 3.6\,arcmin. 

For arrays with longer baselines such as LOFAR or an extended MWA, ionospheric effects will be a much bigger challenge. Second-order spatial TEC derivatives over the array will become significant, leading to refractive shifts in different directions for different baselines. In this case, decorrelation and shape distortions could occur on a regular basis. Arrays with longer baselines will also be able to resolve a greater number density of sources, and cross-matching radii as large as 1--2\,arcmin will be unrealistic. In such cases, a thorough calibration that accounts for the full direction dependence of ionospheric phases may be necessary to achieve the objectives of time-domain astrophysics. Fortunately, for a compact array such as the MWA, ionospheric effects appear not to adversely affect the ability to conduct these types of studies.

Our analysis does not include the effects of phase calibration errors, which manifest as a global shift of sources with respect to the underlying coordinate grid. We removed these by measuring angular displacements with respect to time-averaged positions, which isolates only the short-term fluctuations. However, the global shifts induced by calibration errors can be comparable to those resulting from ionospheric refraction, and present an independent source of difficulty for cross-matching.

We find that short-term amplitude variations contain a component, separate from thermal noise and sidelobe confusion, that is described by a modulation index of $\mu \sim$ 1--3\%. This is an empirical upper bound on ionospheric scintillation, and appears to be more important than thermal noise/sidelobe confusion for sources brighter than about 1\,Jy for typical MWA sensitivities. This component may possibly arise from effects other than scintillation, but we do not pursue the investigation here. We note that ionospheric activity, characterised either by $\langle s \rangle$ or $\mu$, does not exhibit compelling correlations with either the RMS noise or the S/N ratio of sources. Thus, at least under the observing conditions of the datasets used in this study, the ionosphere does not appreciably affect the point-source sensitivity of the MWA.

We detected a large-scale, concave-down modulation pattern in the peak flux density of sources as they transited through the FoV. The variation is substantial, of order 10\%, and is most probably related to an incomplete understanding of the primary beam gain pattern rather than an ionospheric effect (see arguments in \S\ref{sec:fluxinterpretation}). In either case, it does not represent intrinsic variability and its ubiquity is therefore concerning from the point of view of time-domain analyses. Obtaining a better understanding of the characteristics of the MWA primary beam is an area of considerable ongoing work, from both theoretical and empirical viewpoints, described in part by \citet{Sutinjo2014} and \citet{Neben2015_mn2e}.

Finally, we caution that this is not an exhaustive study of all data collected by the MWA. The data analysed in this work all happened to be obtained under quiet geomagnetic conditions, where the $K_p$ index (quantifying global fluctuations of the Earth's magnetic field) was 2 or less. Our results therefore pertain mainly to quiet-time effects of the ionosphere. Data have so far not been obtained for MWATS under significantly disturbed conditions, and so the nature of possible worst-case scenarios, including the incidence rate of scintillation events that may be associated with storm-time activity, remains to be established.

\section{Conclusions}\label{sec:conclusions}
We have analysed 20 MWATS datasets encompassing 10 nights of observations under quiet geomagnetic conditions to establish how the ionosphere affects the feasibility of time-domain science with the MWA at 154\,MHz. The two quantities whose statistical behaviour we examined were angular positions and peak flux densities. We found that:
\begin{itemize}
  \item Angular positions fluctuate by $\sim$10--20\,arcsec, these being below the synthesised beamwidth $>$99\% of the time;
  \item There is a consistent preferred direction for angular position offsets, this being NW-SE (geographic), which matches prior reports of anisotropies observed in ionospheric density structures;
  \item An upper bound on the modulation index associated with ionospheric scintillation is $\sim$1--3\%;
  \item There is a persistent, concave-down modulation of the radio light curves over the MWA FoV, but propagation effects have difficulty accounting for this.
\end{itemize}

It appears that the ionosphere does not adversely affect the feasibility of time-domain science with the MWA at 154\,MHz, at least under conditions similar to those examined here. A cross-matching radius of 1--2\,arcmin is sufficient most of the time, but under more severe conditions it may be necessary to increase this to 3--4\,arcmin. Light-curve error bars may be made more realistic by taking into account the existence of an extrinsic source of amplitude fluctuation characterised by a modulation index of several per cent. The occurrence rates of extreme events cannot be established from our data; this demands a broader investigation under a variety of geomagnetic and tropospheric conditions. We intend to pursue this in future work.

\section*{Acknowledgments}
We thank J.-P.~Macquart and J.~Morgan for useful discussions. This scientific work makes use of the Murchison Radio-astronomy Observatory, operated by CSIRO. We acknowledge the Wajarri Yamatji people as the traditional owners of the Observatory site. Support for the MWA comes from the U.S. National Science Foundation (grants AST-0457585, PHY-0835713, CAREER-0847753, AST-0908884 and AST-1412421), the Australian Research Council (LIEF grants LE0775621 and LE0882938), the U.S. Air Force Office of Scientific Research (grant FA9550-0510247), and the Centre for All-sky Astrophysics (an Australian Research Council Centre of Excellence funded by grant CE110001020). Support is also provided by the Smithsonian Astrophysical Observatory, the MIT School of Science, the Raman Research Institute, the Australian National University, and the Victoria University of Wellington (via grant MED-E1799 from the New Zealand Ministry of Economic Development and an IBM Shared University Research Grant). The Australian Federal government provides additional support via the Commonwealth Scientific and Industrial Research Organisation (CSIRO), National Collaborative Research Infrastructure Strategy, Education Investment Fund, and the Australia India Strategic Research Fund, and Astronomy Australia Limited, under contract to Curtin University. We acknowledge the iVEC Petabyte Data Store, the Initiative in Innovative Computing and the CUDA Center for Excellence sponsored by NVIDIA at Harvard University, and the International Centre for Radio Astronomy Research (ICRAR), a Joint Venture of Curtin University and The University of Western Australia, funded by the Western Australian State government.


\end{document}